\documentclass[preprint2]{aastex}
\usepackage{natbib}
\usepackage{ulem}    %added by LRY
\usepackage{graphicx}  %added by LRY
\def\msun{M$_{\odot}$}
\def\rsun{R$_{\odot}$}
\def\lsun{L$_{\odot}$}

\def\sbs{SBS\,1150+599A}
\def\pn{PN\,G135.9+55.9}

\def\it{\sl}
\def\degs{\ifmmode ^{\circ}\else$^{\circ}$\fi}
\def\amin{\ifmmode ^{\prime}\else$^{\prime}$\fi}
\def\asec{\ifmmode ^{\prime\prime}\else$^{\prime\prime}$\fi}
\def\fd{\hbox{$.\!\!^{\rm d}$}}            % Fractions of days
\def\degs{\ifmmode ^{\circ}\else$^{\circ}$\fi}
\def\amin{\ifmmode ^{\prime}\else$^{\prime}$\fi}

\def\eqalign#1{\null\,\vcenter{\openup1\jot \m@th
   \ialign{\strut\hfil$\displaystyle{##}$&$\displaystyle{{}##}$\hfil
   \crcr#1\crcr}}\,}
\usepackage{color}
   
\def\te{$T_{\mathrm {eff}}$}
\newcommand{\nev}{[Ne~{\sc v}]}
\def\sna{SN\,Ia}

\newcommand{\ace}{\mbox {$\alpha_{ce}$}}
\newcommand{\al}{\mbox {$\alpha_{ce}\lambda$}}
\newcommand{\myr}{\mbox {~${\rm M_{\odot}~yr^{-1}}$}}
\newcommand{\pyr}{\mbox {{\rm yr$^{-1}$}}}
\newcommand{\TS}{TS\,01}
\slugcomment{prepared for Ap.J.}

\shorttitle{The Double-Degenerate Nucleus of the Planetary Nebula \TS.}
\shortauthors{Tovmassian et al.}

\begin{document}


\title{ The Double-Degenerate Nucleus of the Planetary Nebula \TS.\\A Close Binary Evolution Showcase.}


\author{{Gagik  Tovmassian}\altaffilmark{1},%}   %%% Fill in author names
%\affil{Instituto de Astronomia, Universidad Nacional Autonoma
%de Mexico, Apdo. Postal 877, Ensenada, Baja California, 22800 Mexico } %%% Fill in author affiliations
%\email{gag@astrosen.unam.mx}
%\author
{ Lev Yungelson}\altaffilmark{2},   %%% Fill in author names
%\affil{Institute of Astronomy of the Russian Academy of Sciences, 48 Pyatniskaya Str., 119017 Moscow, Russia}    %%% Fill in author affiliations
%\email{lry@inasan.ru}
%\author
{ Thomas Rauch} \altaffilmark{3},  %%% Fill in author names
%\affil{Kepler Center for Astro and Particle Physics, Institute 
%for Astronomy und Astrophysics, University of T\"ubingen, Sand 1, 
%72076 T\"ubingen, Germany}
%\email{rauch@astro.uni-tuebingen.de}
{ Valery Suleimanov}\altaffilmark{3}\altaffilmark{*},  %%% Fill in author names
%\affil{Kepler Center for Astro and Particle Physics, Institute 
%for Astronomy und Astrophysics, University of T\"ubingen, Sand 1, 
%72076 T\"ubingen, Germany}
%\altaffiltext{1}{Society of Fellows, Harvard University.}
%\email{suleimanov@astro.uni-tuebingen.de}
%\author
{ Ralf Napiwotzki}\altaffilmark{4},   %%% Fill in author names
%\affil{Centre for Astrophysics Research, University of Hertfordshire, College Lane, HatÞeld 
%AL109AB, UK  }    %%% Fill in author affiliations
%\email{rn@star.herts.ac.uk}
%\author
%\author{
{ Grazyna Stasi\'nska}\altaffilmark{5},  %%% Fill in author names
%\affil{LUTH, Observatoire de Paris, CNRS, Universit\'e Paris Diderot; Place Jules Janssen 92190 Meudon, France}    %%% Fill in author affiliations
%\email{grazyna.stasinska@obspm.fr}
%\author
{ John Tomsick}\altaffilmark{6},   %%% Fill in author names
%\affil{Space Sciences Laboratory, 7 Gauss Way, 
%University of California, Berkeley, CA 94720-7450, USA.}    %%% Fill in author affiliations
%\email{jtomsick@ssl.berkeley.edu}
%\author
{ J\"orn Wilms}\altaffilmark{7},   %%% Fill in author names
%\affil{Dr. Karl Remeis-Observatory, University of Erlangen-Nuremberg, Sternwartstrasse 7, 96049 Bamberg, Germany}    %%% Fill in author affiliations
%\email{joern.wilms@sternwarte.uni-erlangen.de}
%\author
{ Christophe Morisset}\altaffilmark{8},   %%% Fill in author names
%\affil{Instituto de Astronomia, Universidad Nacional Autonoma de Mexico, Apdo. Postal 70264, Mexico D.F., 04510 Mexico}%%% Fill in author affiliations
%\email{morisset@astroscu.unam.mx}
%\author
{ Miriam Pe\~na}\altaffilmark{8},   %%% Fill in author names
%\affil{Instituto de Astronomia, Universidad Nacional Autonoma de Mexico, Apdo. Postal 70264, Mexico D.F., 04510 Mexico}%%% Fill in author affiliations
%\email{miriam@astroscu.unam.mx}
%\and
%\author
{ Michael G. Richer}\altaffilmark{1}}   %%% Fill in author names
%\affil{Instituto de Astronomia, Universidad Nacional Autonoma
%de Mexico, Apdo. Postal 877, Ensenada, Baja California, 22800 Mexico}%%% Fill in author affiliations
%\email{richer@astrosen.unam.mx}


\altaffiltext{1}{Instituto de Astronomia, Universidad Nacional Autonoma
de Mexico, Apdo. Postal 877, Ensenada, Baja California, 22800 Mexico.}
\altaffiltext{2}{Institute of Astronomy of the Russian Academy of Sciences, 48 Pyatnitskaya Str., 119017 Moscow, Russia.}
\altaffiltext{3}{Institute for Astronomy and Astrophysics, Kepler Center for Astro and Particle Physics, 
Eberhard Karls University, Sand 1, 72076 T\"ubingen, Germany. }
\altaffiltext{4}{Centre for Astrophysics Research, University of Hertfordshire, College Lane, HatÞeld 
AL109AB, UK.}
\altaffiltext{5}{LUTH, Observatoire de Paris, CNRS, Universit\'e Paris Diderot; Place Jules Janssen 92190 Meudon, France.}
\altaffiltext{6}{Space Sciences Laboratory, 7 Gauss Way, 
University of California, Berkeley, CA 94720-7450, USA.}
\altaffiltext{7}{Dr. Karl Remeis-Observatory, University of Erlangen-Nuremberg, Sternwartstrasse 7, 96049 Bamberg, Germany.}
\altaffiltext{8}{Instituto de Astronomia, Universidad Nacional Autonoma de Mexico, Apdo. Postal 70264, Mexico D.F., 04510 Mexico.}
%\altaffiltext{9}{Department of Astrophysics, Radboud University Nijmegen, Heyendaalseweg 135,
%NL-6525 AJ Nijmegen, Netherlands.}
\altaffiltext{*}{Kazan State   University, Russia.}

\begin{abstract} 
We present a detailed investigation  of  \sbs, a close binary star 
hosted by the planetary nebula   PN\,G135.9+55.9 \citep[\TS, 
][]{paperI}. 
The nebula, located in the Galactic halo,  is  the most oxygen-poor 
one  known to date and is the only one known to harbor a double 
degenerate core. We present XMM-{\sl Newton}  observations of this 
object, which allowed the detection of  the previously invisible 
component of the binary core, whose existence was inferred so far 
only from radial velocity and photometric variations. 
The parameters of the binary system were deduced from a wealth of 
information via three independent routes using the spectral energy 
distribution (from the infrared to X-rays), the light and radial 
velocity curves, and a detailed model atmosphere fitting of the 
stellar absorption features of the optical/UV component. We find 
that the {\sl cool} component must have a mass of 
$0.54\pm0.2$\,\msun, an average effective temperature, T$_{\mathrm 
{eff}}$, of $58\,000\pm3\,000$\,K, a mean radius of 
$0.43\pm0.3$\,\rsun, a gravity $\log g=5.0\pm0.3$, and that it nearly 
fills its Roche lobe.  Its  surface elemental abundances are found to 
be: 12 + log He/H = 10.95 $\pm$0.04\,dex,  12 + log C/H = 
7.20$\pm$0.3\,dex, 12 + log N/H $<$ 6.92 and 12 + log O/H $<$ 6.80, 
in overall agreement with the chemical composition of the planetary 
nebula. 
The  {\sl hot} component has T$_{\mathrm {eff}}$ =  160--180\,kK, a 
luminosity of about $\sim 10^4$\lsun\ and a radius slightly larger 
than that of a white dwarf.   It is probably bloated and heated as a 
result of intense accretion and  nuclear burning on its surface  in the past.  The total mass of 
the binary system is very close to Chandrasekhar limit.  This makes \TS\ one of the best   type Ia 
supernova  progenitor candidates. We propose two possible scenarios 
for the evolution of the system up to its present stage. 

\end{abstract}




\section{Introduction}

\objectname{\sbs} was identified as a planetary nebula  (PN) in \cite{2001A&A...370..456T} and subsequently designated as \objectname{\pn}.  More recently, we refer to this  object  as \TS\ \citep{paperI}. The object has unusually few spectral lines for a PN  and is renown for its extremely low oxygen content \citep{2001A&A...370..456T, 2002AJ....124.3340J, 2005A&A...430..187P,2005IAUS..228..323S, paperI}. It is located above the Galactic plane at a distance of at least a dozen kpc, which places it among  a handful of known halo PNe. Direct images obtained on the ground  \citep{2002A&A...395..929R,2002AJ....124.3340J}, and most recently by HST \citep{2005AIPC..804..173N, paperI} confirm its PN identification. The observed expansion velocity of the nebula \citep{2003A&A...410..911R} is typical of  PNe.  But another outstanding feature of this PN is that it harbors a close binary system \citep{2004ApJ...616..485T}, revealed serendipitously by the displacement of stellar lines with respect to nebular lines within a single observing night.  Since only one component of the binary could be observed  in the optical and UV, it was suggested that the visible  component has a temperature of 110\,000--120\,000\,K.  The lower limit is the minimum effective temperature  needed to produce the  observed  [Ne\,V] nebular emission line, while the upper limit was deduced from the slope of the continuum   \citep{2004ApJ...616..485T}. There was an ambiguity in the determination of the orbital period, although  it was clear that the nucleus is a close binary with a period less than 4 hours. The high temperature, coupled with high log\,$g$, determined  from the profiles of absorption lines, led all studies prior to \citet{paperI} to assume that the observed optical/UV component was  the central star of the planetary nebula, i.e. the post-AGB star that lost its envelope and was the source of its ionization.
\citet{2005A&A...430..187P} suggested that, if the ionizing star were even hotter, the deduced oxygen abundance could be increased to a more  common level for oxygen-poor PNe. However, a higher temperature would have required a higher reddening to match the observed continuum slope, and \citet{2004ApJ...616..485T} had already used a higher extinction than would normally be estimated for the direction of \sbs\  in order to justify a temperature of  120\,000\,K  .

Next, we obtained photometric light curves of the binary core of \sbs\ \citep{2005AIPC..804..173N}. The orbital period of the system turned out to be  3.92\,hr and, to explain the double-humped shape of the light curve, we were led to invoke a Roche lobe-filling optical/UV component. It was observed that the depths of the minima in the light curve are uneven, an effect known to occur when the visible component is irradiated by  a hotter  (more energetic)  source. 
The orbital dynamics required that this invisible component be another compact object  of at least 0.85\,\msun\  \citep{2005AIPC..804..173N}. \citet{2002AJ....124.3340J}  pointed out the possibility that \objectname{\sbs} may be associated with the X-ray source \objectname{1RXS\,J115327.2+593959}.   To detect the invisible source of irradiation and reveal the other component of the close binary, we observed it with the XMM-{\sl Newton}  X-ray observatory.  We also conducted new optical spectroscopic observations of the object with the Gemini-North telescope to improve our knowledge of radial velocities of the optical/UV component of the binary and to better fit photospheric line profiles with atmospheric models. We also  used the publicly available HST STIS observations of the object in the UV to bridge the optical and X-ray observations discussed here.  

The ionization state and chemical composition of the planetary nebula are analyzed in a companion paper \citep{paperI}, while here we present a multifaceted analysis and modeling of the binary system. We analyze the history and the future  of the stellar system in the light of evolutionary models for close binary stars.

 In Section 2, we present  our new  observations;  in Section 3 we determine the physical parameters of the binary; 
 in Section 4  we discuss the evolution of the object from the early stages, when it was a wide system comprised of main sequence stars, to the latest stage of a  merging of two white dwarfs (WD) with possible type Ia supernova outcome; and in  Section  5 we summarize our main results.

\section{Observations.}

%% In a manner similar to \objectname authors can provide links to dataset
%% hosted at participating data centers via the \dataset{} command.  The
%% second curly bracket argument is printed in the text while the first
%% parentheses argument serves as the valid data set identifier.  Large
%% lists of data set are best provided in a table (see Table 3 for an example).
%% Valid data set identifiers should be obtained from the data center that
%% is currently hosting the data.
%%
%% Note that AASTeX interprets everything between the curly braces in the 
%% macro as regular text, so any special characters, e.g. "#" or "_," must be 
%% preceded by a backslash. Otherwise, you will get a LaTeX error when you 
%% compile your manuscript.  Special characters do not 
%% need to be escaped in the optional, square-bracket argument.

%\objectname{\TS} has now been observed from the far IR to  X-rays in all possible wavelengths.
%In the optical/UV, we mostly see  the continuum emission of a hot star upon which are superposed the emission lines from the nebula.  The stellar continuum displays absorption %lines as well as interstellar absorption features (mostly in the  UV).

\subsection {Optical observations.}

A spectrum of \TS, with an ample coverage of  3800--9200\,\AA\  is available in the Sloan Digital Sky Survey (SDSS\footnote{http://www.sdss.org}). It was taken on 2002 May 17 (spSpec-52411-0953-160). We used the newly calibrated spectrum that appears in the SDSS Data Release 7.  This spectrum provides  probably the best flux calibration, in a perfect agreement with HST spectra (see below).  

The stellar absorption lines  of the Balmer series and of He\,II are difficult to detect due to their shallowness, the faintness of the object (V$\sim 18^{\mathrm m}$), and the presence of very intense emission lines from the nebula. Only 7 spectra with measurable absorption features were available from our CFHT observations \citep{2004ApJ...616..485T}. The short orbital period and consequent line smearing by the long exposures, and the necessity of relatively high resolution to effectively disentangle emission lines from absorption indicated the need for observations with a larger telescope.

\begin{figure}[t]
\includegraphics[width=7.8cm,bb = 0 5 380 380, clip=]{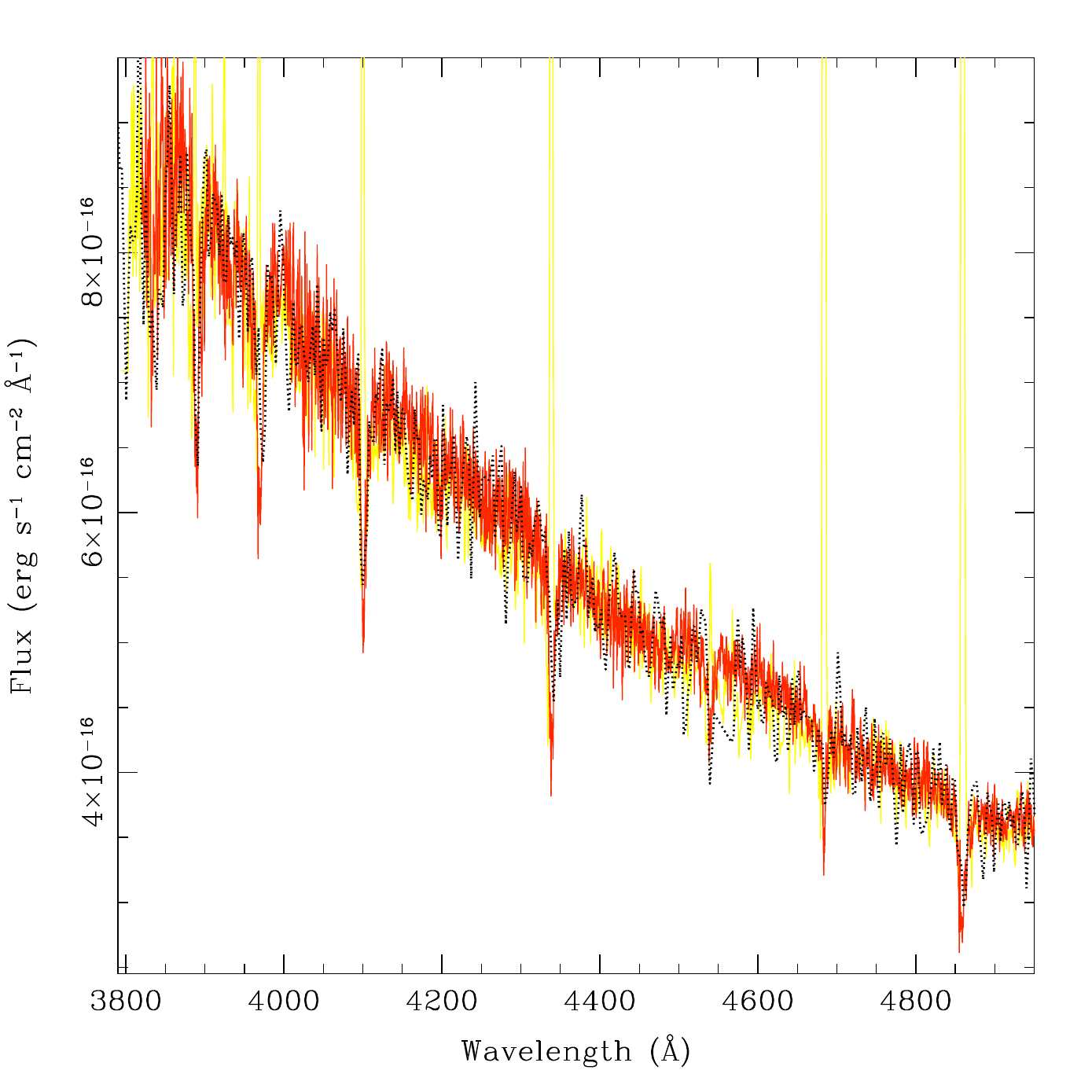}
\caption{The stellar spectrum of \TS\ obtained at Gemini North (solid red line). The spectrum is a combination of 12 individual spectra, observed at different orbital phases, and smoothed with a 13 pixel boxcar.  Before combining, the individual spectra were corrected for orbital motion.  The SDSS spectrum, containing nebular lines, is plotted in yellow.  The black  dotted line is the single HST spectrum obtained for the same spectral range (see the text).}\label{geminispec}
\end{figure}

We proposed to observe \objectname{\TS} for a total of 16 hours, covering four orbital periods, with Gemini-North telescope. 
The observations were scheduled for service mode in semester 2006A, but only 20\% were completed. The available observations were performed in two sets: on 2006 May 16 UT, 8 spectra were obtained, and, on 2006 Jun 09, four more were added. The weather conditions during  the observations were not ideal. %{\color{red}\sout{ particularly during the first set of observations}}. {\color{red} \footnote{
The exposure times were 700\,s, so, in total, only about 3/4 of the orbital period was covered with a resolution in phase of 5\%.
The GMOS spectrograph was used with the B1200+G5301 grating, leading to effective spectral resolution of 1.65\,\AA\ (FWHM) and a spectral coverage of 3800--5000\,\AA.  We observed the Balmer series from H$_\beta$\   to the highest  members (H$_9$ ) %{\color{red}\sout{, around which the contribution of the nebular emission lines is the smallest}}. 
This spectral range also includes the He\,{\sc ii} 4686\,\AA\ line, detectable in both emission and absorption.   Auxiliary images (biases, flatfields, arcs) were used to reduce the data using the procedures in {\sl gemini} package within IRAF\footnote{Copyright(c) 1986 Association of Universities for Research in Astronomy Inc.} and prescriptions provided by Gemini staff and fellow observers\footnote{http://www.astro.caltech.edu/\~kelle/gmos/gemini\_reduction.html}.  
 The standard star PG\,1545+035, observed with the same instrumental configuration on 2006 Aug 30 in apparently better conditions, was used in an attempt at flux calibration. However, the result of this calibration was not satisfactory.  Instead, we used the spectrum of \TS\ available in the Sloan Digital Sky Survey (SDSS) data base to correct the continuum. %This correction was primarily done in order to be able to combine all spectra and obtain a high S/N spectrum of the object. Since we were interested in the profiles of the absorption lines, the details of the flux calibration should not affect our results.
The {\sl Gemini} spectra were corrected for orbital radial velocity shifts using the orbital solution described below and then co-added. Combining the 12 radial velocity (RV) shifted spectra allowed us to improve the profiles of stellar absorption lines and to get rid of nebular emission lines. The resulting spectrum, after 13 pixel boxcar smoothing, is presented in Fig.\,\ref{geminispec}. 

Previous observations of \TS\ are used here to analyze the nature of the stellar core. In addition to the above mentioned 7 CFHT spectra of lower spectral resolution, they include   multiband photometric observations, briefly presented in \citet{2005AIPC..804..173N}. They were obtained on two consecutive nights with the 2.2\,m telescope at Calar Alto and the BUSCA CCD camera system that allows simultaneous direct imaging in four colors. 
The differential  photometry was performed using comparison stars in the field of view.  These photometric data were complemented by CCD photometry in the $V$ filter obtained with the 2.1\,m telescope at the Observatorio Astron\'omico Nacional in the Sierra de San Pedro M\'artir (OAN SPM) on 2004 Apr 09.  Additional photometric data were provided by the optical monitoring instrument (OM) on board XMM during the X-ray observations.  Optical and UV data from direct images in the optical range as well as integrated flux from spectroscopic observations were also incorporated into the time series. For the time series analysis, the photometric data from the different wavelengths and bandpasses  were normalized to a common mean value and combined.  

\begin{figure}[ht]
\includegraphics[width=7.8cm,bb = 20 50 720 550, clip=]{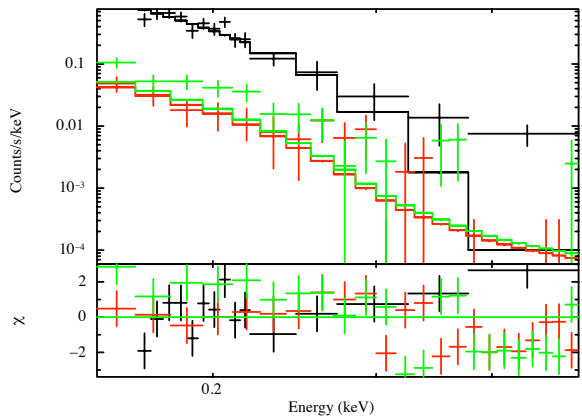}
\caption{The source spectrum in all three detectors, PN, MOS1, and MOS2 (black, red and green crosses respectively) are fitted with the same black body model (lines of corresponding colors). The photons are concentrated in the lowest 0.1-- 0.3 keV range.  There is  general consistency between all three EPIC detectors at energies below 300 eV. In the 0.4 -- 0.5 keV bin, the counts start to differ, but the statistical significance of this is very low taking into account low count rates and calibration uncertainties. }
\label{fig2}
\end{figure}

\subsection{Ultraviolet observations.}

The first UV data for \TS\ were obtained in the far-UV using the {\sl Far Ultraviolet Spectroscopic Explorer} satellite (FUSE). Details of these observations and their results are provided in  \citet{2004ApJ...616..485T}. 

Later,  observations  in the near-UV were obtained with the Hubble Space Telescope (HST)  (Obs. ID 9466).  In 2003 May, HST obtained spectra of \TS\ with the Space Telescope Imaging Spectrograph (STIS) in FUV, NUV, and CCD  modes to cover the entire UV and optical. 
Continuous (uninterrupted by the Earth occultations)  exposures with the G140L  and G230L  gratings were acquired. Five spectra with each grating and with individual exposure times of 4675\,s and 2850\,s, respectively, were acquired and combined to produce the final spectrum.  One 600\,s exposure was taken of the UV-optical  spectrum  (G430L grating)  to connect the UV data with the optical.  This last spectrum, as a result of its short exposure time, fails to reveal relatively weak, though important, emission lines in the optical UV, but provides a decent stellar spectrum that overlaps nicely with the NUV and SDSS spectra.  This spectrum is also plotted in Fig.\,\ref{geminispec} and is in  very good agreement regarding absorption features.   
The object was also observed in 2003 June with High Resolution Camera (HRC) of the Advanced Camera for Surveys (ACS) in the F334N \& F658N  filters to obtain images in the strong nebular lines of [Ne\,{\small V}] and H$_\alpha$, respectively. 

The pipeline-reduced STIS spectra were utilized to extract the stellar continuum. % Here, we use the HST STIS data to refine our temperature estimates of the stellar core of \TS.  
The integrated fluxes of individual exposures in UV range were also summed to produce a light curve in the UV range.

\subsection{X-ray observations.}

\TS\ was observed with XMM-{\sl Newton} (obs-ID 0404220101) on 2006 Nov 01--02 (revolution 1263) in a continuous  27\,ks exposure. The X-ray-counting EPIC instruments were operated with the thin filter in the PN small window mode and full window for the MOS detectors. The object was too faint for EPIC-RGS detectors. The optical monitor (OM) instrument on board XMM-{\sl Newton} took 16 images in a $B$ filter, each of 22 minutes duration.  No pile-up was detected in either of the EPIC detectors.  Background photon flares  were detected during only 35-40\% of exposure, mostly towards its end. The 7.5 hour exposure is just shy of two orbital periods ($2\times3.9$ hours) of the binary system.  The data from the first orbit were completely free from background flaring effects.  The observed mean source count rates were $0.033\pm0.002$\ in the PN, $0.0025\pm0.0004$\  and $0.0053\pm0.0005  \,{\rm cnts\, s}^{-1}$, in the MOS1 and MOS2, respectively.

The data were reduced using XMM-SAS (version 9.0).  
%To extract light curves and spectra in the PN and MOS detectors, we used macros and procedures developed by J. Wilms. The macros and their description are provided at his website\footnote{http://astro.uni-tuebingen.de/$\sim$wilms/research/analysis/}.   
For the MOS detector, the source and background photons were extracted  from a circular aperture and surrounding annulus correspondingly. For the PN detector, we tried subtraction of background from 2 different circles near  the source, since the small window did not allow to use a large annulus. We found no substantial differences in background removal from different areas. Events with detection patterns of up to quadruples were selected.  
%No big changes are noted between the old and new reductions, although the calibration has been improved significantly from version 7 to 9. The data analysis and interpretation throughout the paper are based on  the new reduction.
 
Background-subtracted spectra in the three EPIC detectors and a single blackbody model corresponding to each detector are shown in Fig.\ref{fig2}.  Estimates of the total galactic  H\,{\sc i} column density varies  from $N_{\mathrm H}=1.8\times10^{20} $ \citep{2005A&A...440..775K} to $N_{\mathrm H}=1.53\times10^{20} \rm{cm}^{-2}$  \citep{1990ARA&A..28..215D}. The X-ray spectral analysis, with the column density fixed to the mean value determined 
for the direction of \TS\   ($N_{\mathrm H}= 1.6\times10^{20} \rm{cm}^{-2}$), gives a best fit for  $kT = 17$\,eV (T$\approx195\,000$ K).  
Based upon the XSPEC \citep{1996ASPC..101...17A} modeling, the 90\%  confidence region spans $kT = 12.7-18.0$ eV. The source is extremely soft and emits only in the narrow range spanning 100 -- 300 eV.  This range is notorious for its unreliable calibration (e.g. \citet{2009A&A...496..879M} for the latest evaluation) and routinely excluded by observers. However there is a good agreement between flux in the PN and MOS detectors  for 0.1--0.3 keV (Fig. \ref{fig2}).  
Although the PN detector suggests the possibility that there is   emission in excess of the blackbody in the 0.4-0.5 keV bin, the excess is not   confirmed by either of the MOS detectors, making it unlikely that it is a real  spectral feature. The analysis of ROSAT RASS archival data reveals that the source was poorly covered  and the background is uneven, making spectral fitting useless, though it does confirm the extreme softness of the source. 

\begin{figure}[ht!]
\includegraphics[width=7.8cm,bb = 0 5 380 400, clip=]{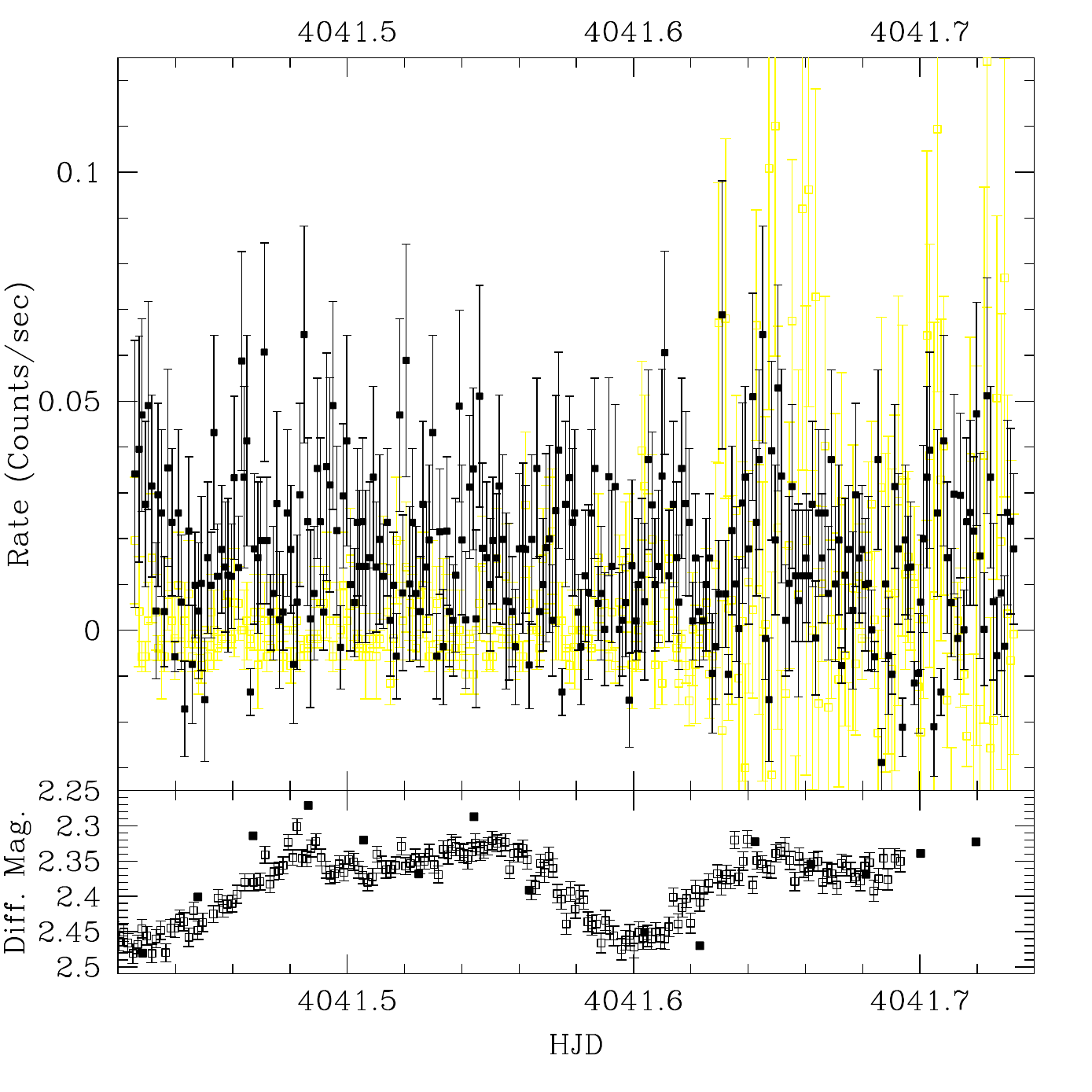}
\caption{The X-ray light curve  of the source in the PN detector (top panel). The filled symbols represent the light curve in 0.1--0.3 keV range, while the open symbols (yellow) are the counts in the 0.3--10.0 keV band.  The count rate above 0.3 keV is practically zero, except during the final part of the exposure, where they are dominated by background flaring. In the extreme soft  band, 0.1--0.3 keV, the count rate is approximately constant (0.033 cnts/sec) for the entire duration of the observation.  The bottom panel displays the optical light curve for comparison. Black points are OM measurements on board of {\sl XMM-Newton} at the time of the observations whereas the open boxes are $V$ light curve obtained at SPM (2.5 years earlier; shifted in time according to phases).}
\label{Xlc}
\end{figure}

In Fig. \ref{Xlc}, the light curve of the source in the PN detector is presented extracted in two energy bands, 0.1-0.3 keV and 0.3-10 keV.  This light curve demonstrates that practically all photons from the source are emitted in the narrow soft band, that the background flaring occurs mostly in the last quarter of the 27\,ks exposure, and that the  flares do not affect the soft band, where the source emits, but are rather   strong in higher energies, confirming their nature as  background.  A similar picture emerges from the MOS detectors. However, we have chosen a conservative approach and excluded all episodes when the count-rate  exceeded 0.1 counts/s  from the  analysis of the source in all three detectors.
 
The X-ray light curve of the source in the PN detector shows some flickering but no definite periodic variability. There  is no correlation between the  power spectra obtained for the three different detectors and the optical light curve. % {\bf The flickering is believed to be a sign of some continuing accretion in the system \citep{2010MNRAS.401..121P}.}
 
The OM measurements in the B filter  from the pipeline reduction were simply transformed into a magnitude scale by taking  the logarithm  and shifted   to the same average value as the ground-based optical differential photometry.  In this way, they were used to identify the precise orbital phasing for the X-ray data.

\begin{figure}[ht]
\epsscale{1.2}
\includegraphics[width=8.cm,bb = 22 20 450 320, clip=]{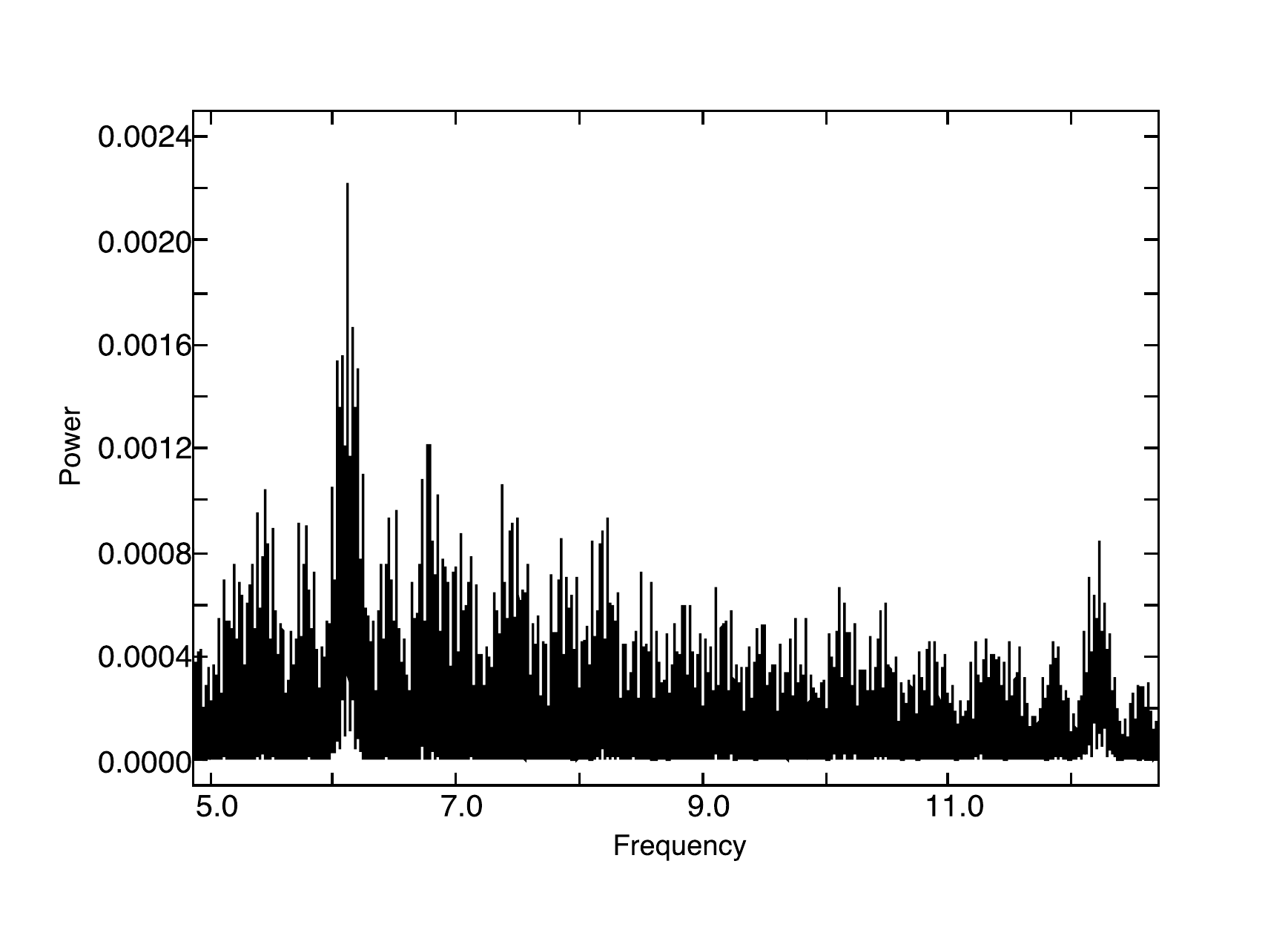}
%\plotone{tovmassian_fig4n}
\caption{The power spectrum of all photometric data. The strongest, single peaked maximum corresponds to the orbital period, at 6.1159\,day$^{-1}$. The first harmonic, at $\approx 12$\,days$^{-1}$, is also prominent in the power spectrum because of the double-humped nature of the light curve. 
}
\label{powspec}
\end{figure}

\begin{figure}[ht!]
\includegraphics[width=8.cm,bb = 3 0 400 400, clip=]{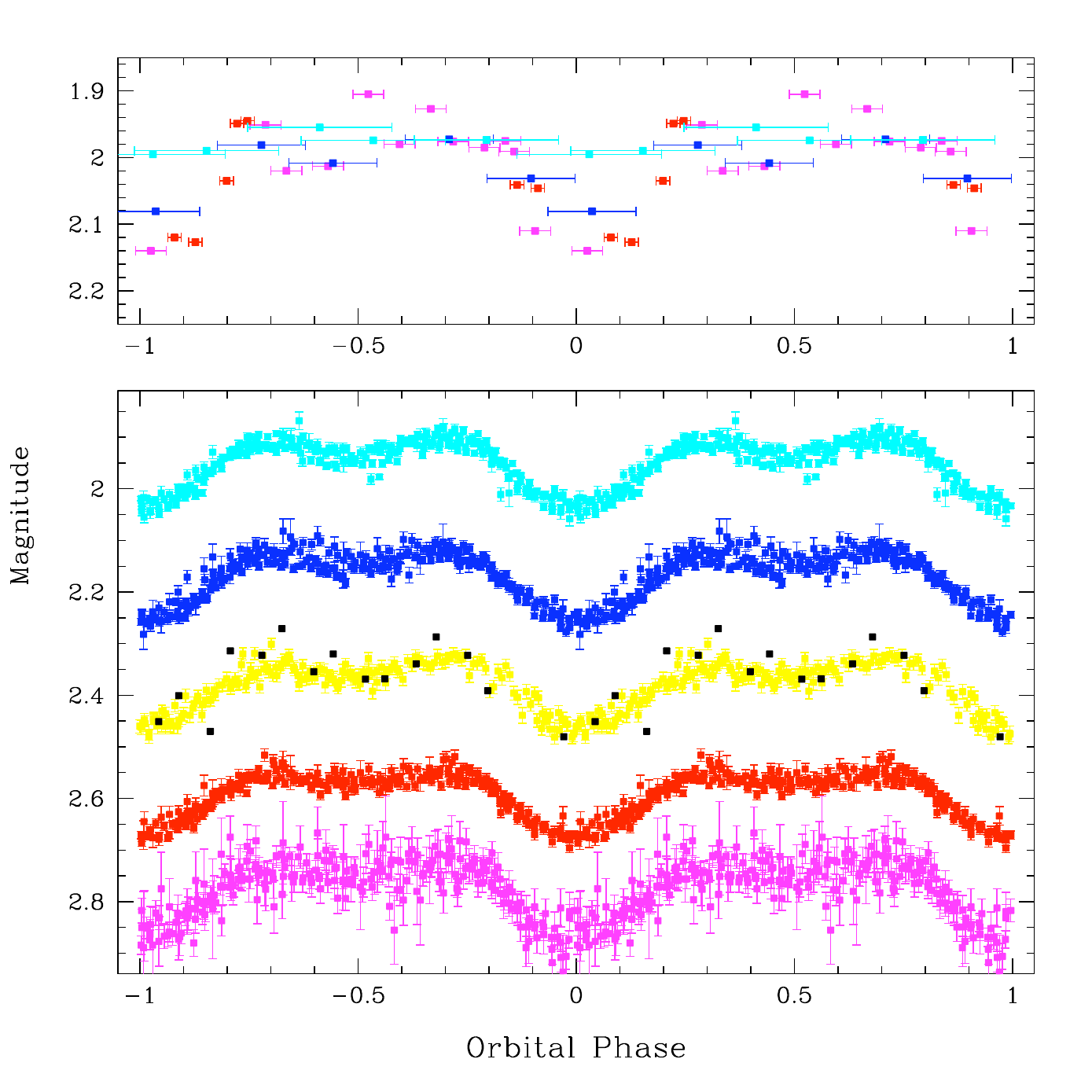}
\caption{The light curve of \TS\ folded with the orbital period. The lower panel presents optical light curves. From top to to bottom are: BUSCA UV and $B$, Johnson $V$ from SPM, XMM OM (black points), and finally BUSCA R and NIR.
The upper panel presents HST data:  H$_\alpha$ (red), \nev\  (magenta), the integrated spectral flux in the NUV (dark blue) and FUV (cyan). The horizontal error bars in the upper panel denote the exposure time converted to orbital phase. }\label{sbslc}
\end{figure}

\section{Physical parameters of the close binary.}
\subsection{Orbital Period.}

%\begin{figure}[ht!]
%\includegraphics[width=3.75cm,bb = 3 0 560 780, clip=]{tovmassian_fig6a}
%\includegraphics[width=3.75cm,bb = 3 0 560 780, clip=]{tovmassian_fig6b}
%\caption{Examples of absorption line profiles fitted using the FITSB procedure. The emission component from the nebula was removed by hand from the individual spectra before fitting.}
% \label{fitsb}
%\end{figure}

\begin{figure}[hb!]
\includegraphics[width=8.cm,bb = 3 0 400 400, clip=]{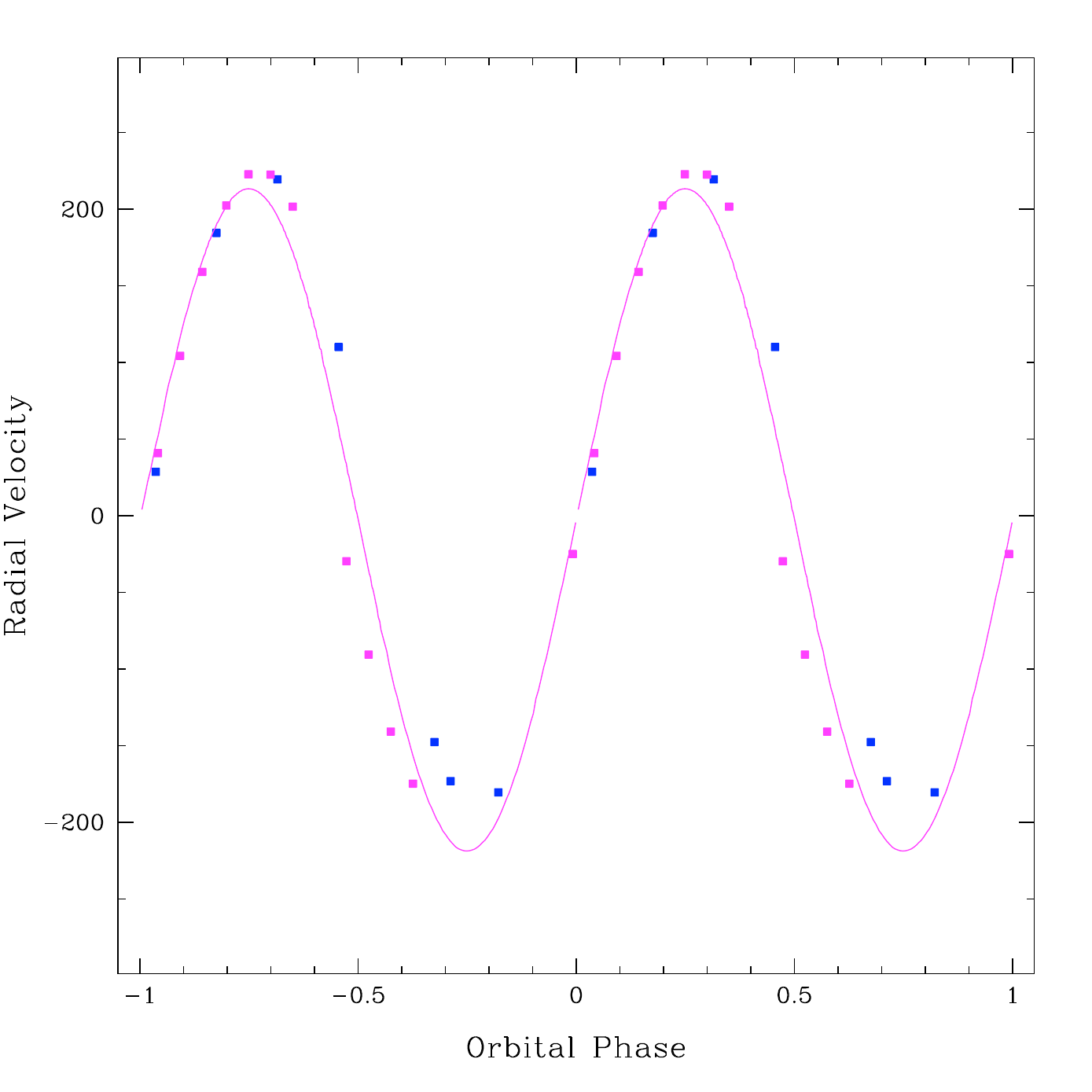}
\caption{The radial velocity curve of \TS\ folded with the orbital period.  The points marked in magenta are measurements from {\sl Gemini} spectra. The CFHT points are presented in blue.  The latter were measured anew with the same parameters deduced from the line  profile fitting by FITSB2.  The $sin$ fit to the data presented here is practically indistinguishable from RV curve calculated by {\sl Nightfall} (see below).}
 \label{rv}
\end{figure}

The orbital parameters were roughly determined  with the discovery of the binarity \citep{2004ApJ...616..485T}. However the precise orbital period can be determined only with the accumulation of enough data over a sufficiently long period of time. 
%The proximity of the orbital period to 4 hours makes the task even more complicated.  
At the moment, we have uneven temporal coverage  spanning about 1000 days, combining multicolor photometry from the ground, optical  monitoring  on board of XMM-{\sl Newton} and narrow-band photometry from HST in the optical range. We also considered the integrated flux in narrow bands from UV spectra obtained with HST.  All of the photometric data 
were shifted to an arbitrary mean magnitude and analyzed  for the presence of a period using the discrete Fourier transform (DFT) method.  
%The light curve is double-humped, which increases the power in the first harmonic of the main frequency corresponding to the orbital period, and worsens the aliasing of the power spectrum. However, the {\sl Clean} procedure \citep{1975Ap&SS..36..137D} does a remarkable job of suppressing  aliases and unambiguously picking the right period. 
The power spectrum is presented in Fig.\,\ref{powspec} with its strongest peak at 6.1159 cycles/day, the frequency corresponding to the orbital period, and its $2f$ alias at 12.23 cycles/day. The resulting ephemeris is
  
$$ {\rm HJD}= 2452760.6756(5) + 0\fd163508(3) \times {\rm E},  $$ 

where the zero point $ \rm T_{\rm 0}$  corresponds to the deeper minimum in the light curve. The light curves of \TS\ folded with the estimated 3.924\,h orbital period   are presented in Fig.\,\ref{sbslc} in  different bands.  In the bottom panel multi color photometry from Calar Alto is plotted combined with $V$ band data from OAN SPM and OM-XMM. In the top panel the measurements from a variety of HST detectors are displayed. Even though the HST data in the optical narrow filters F334N \& F658N include large contributions from the nebular emission lines, they nevertheless show variability of the stellar core with a similar  amplitude  as in the broad-band filters. % In the UV, the photometry was obtained by  integrating the  flux from individual spectra over a wide wavelength range covering the entire spectrum.  
The UV light curves have the same double-humped shape as their optical counterparts, but the amplitude decreases as the wavelength moves further into UV.  The far UV observations with the G140L grating have exposures that are almost  a quarter  of the orbital period, so orbital smearing is severe.  Degrading the optical light curves to a similar time resolution shows that the small amplitude in  the far UV  light curve is the result of smearing rather than an actual change in the amplitude of the variation.

The possible interpretation of the double humped light curve was briefly discussed in \cite{2005AIPC..804..173N}. With the X-ray observations in hand we are now confident that the double hump is a result of the surface projection of the ellipsoidal  binary component that fills its Roche lobe and orbits its more massive companion on a relatively high-inclination orbital plane  (to the line of sight), coupled with the effect of gravitational darkening.  The difference in minima dips, on the other hand, is the consequence of the heating of the surface of the Roche lobe-filling star that faces a hotter, but much more compact companion. This phenomenon is  often observed in compact binaries. It is also referred to as reflection \citep{1971ApJ...166..605W}. In what follows, we  qualify by {\sl cool} or {\sl optical} the  Roche lobe-filling component, since it is the main contributor of light in the optical range, and by {\sl hot} or {\sl X} the hotter  component which irradiates its  cooler companion. We refrain from the usual wording of primary and secondary components in this paper, since, as we later discuss, the roles of primary and secondary changed during the evolution of this system. %In what follows, we will refer to the Roche lobe-filling component as the {\sl cool} component or the {\sl optical} component, since it is the main contributor of light in the optical range, and we will call the {\sl hot} or the {\sl X} component, the hotter and compact  component of the binary system that is responsible for the irradiation of its  cooler companion. We refrain from the usual notation of components as primary and secondary in this paper, since, as we will later discuss, the roles of primary and secondary changed at different points of this system's evolution.

The optical spectra are too sparse to  determine the orbital period independently, but they  cover almost all orbital phases. The orbital period estimate from the photometry is good enough to calculate the orbital phases for the spectra.  Therefore, by measuring the radial velocities of the absorption features in each spectrum, we are able to construct the radial velocity curve. We used the FITSB2 procedure to measure the RV of absorption features. FITSB2 performs a simultaneous fit of the spectra covering different orbital phases, i.e., all available information is combined into the parameter determination procedure \citep{2004ASPC..318..402N}.  The fit results are the stellar parameters as well as the RVs \citep{2004ASPC..318..402N}.  We followed the same procedure as  \cite{2004ApJ...616..485T} adding the new {\sl Gemini} spectra.  However, here, we sought solutions with relatively low temperatures, because we now had a better idea of the temperature composition of the binary  \citep{2007arXiv0709.4016T}. The best fits were achieved with T$_{\mathrm  {eff}}=60\,000\pm5\,000$\,K,  $\log g= 5.17 \pm 0.07$  %(held fixed). 
% An example of the fits to the Balmer lines in the Gemini spectra by FITSB2 is presented in Fig.\,\ref{fitsb}.

The radial velocity (RV) curve presented in Fig.\,\ref{rv} is fitted with a simple sinusoid. The phase zero corresponding to the -/+ crossing of the RV curve coincides with the deeper minimum in the light curve. It confirms our interpretation that, at phase 0.0, the {\sl optical} component is seen in conjunction from behind, turning to the observer its smallest projected area and coolest surface temperature.  At phase 0.5, the {\sl optical} component is seen with  the same projected area, but presents the face with the highest temperature, as result of the intense heating from the {\sl X} component. %   set crvsec=41.8+202.6*sin(2*PI*(tt/Porb+0.998943))
A formal orbital solution  leads to semi-amplitude of RV K$_{\mathrm {cool}}$ = $216\pm10$ km\,s$^{-1}$,  and systematic velocity of the system $\gamma= 0\pm 12$ km\,s$^{-1}$ relative to the nebular emission lines.  % The rms deviation about the fit is 31.0 km\,s$^{-1}$, which is high, mostly because of the poor quality of the data around phase 0.75.

The correct determination of phases and RVs allowed us to combine all 12 {\sl Gemini} spectra, eliminating the nebular emission line components, and delineating the absorption profiles with increased signal-to-noise.  The co-added spectrum is shown in  Fig.\,\ref{geminispec}. In combination with the UV stellar spectra from the HST observations, where the stellar component is easily separated from the nebular one due to the high spatial resolution, the coadded spectrum allows us to perform a model atmosphere analysis of the cool component  (see Sect. \ref{sec: temp-radii}. %As the first step we need their temperatures to move forward.

\subsection {Temperatures and Radii.}
\label{sec: temp-radii}

%\citet{2004ApJ...616..485T} concluded that the ionizing star of \TS\ had an effective temperature of $\approx115$\,kK.  By then, it was clear that the nucleus of \TS\ is a binary, so they assumed that it was the hottest star in the binary system that ejected the observed nebula and was entirely responsible for ionizing it.  This was a natural assumption, since, until now, all of the known binary cores of planetary nebulae contain a cool, un-evolved companion.  %It was hard to imagine that the binary might contain an even hotter star, but 
\TS\ is clearly the first planetary nebula known to contain a double-degenerate binary  \citep{2008AJ....136..323D}.  The first indication that the previous interpretations \citep{2004ApJ...616..485T,2005A&A...430..187P}  of the central star of \TS\  may be incorrect  came from the shape of its light curve \citep{2005AIPC..804..173N}.  Now, armed with the X-ray data, we know that even a 130 kK  star can not produce the observed X-ray flux and that an additional component is required to explain the observed spectral energy distribution (SED).

%Our quest for the  binary component parameters was conducted along three independent paths:  (a) simultaneous fitting of the SED by two black bodies, (b) simultaneous modeling of light and RV curves, and (c) fitting stellar atmosphere models to the spectra of the object.   As a first approximation we use a black body  fitting to the SED. 
We analyze the SED, fitting it with a blackbody as a first approximation.
The actual atmosphere can be significantly different from a black body at wavelengths shorter than 900\,\AA, but as starting point a black body gives us a good idea of what we are dealing with. We will discuss deviations from blackbody  later in the paper. % \citet{2004ApJ...616..485T} had only optical and far-UV data from the FUSE. 
%The interstellar absorption law in the far-UV range is  poorly known  and is based on extrapolation \citep{1999PASP..111...63F}, so \citet{2004ApJ...616..485T} used a higher reddening $E(B-V)$ than expected in direction of \TS\ \citep[e.g.,][]{1998ApJ...500..525S} to fit the  data with a  T$_{\mathrm {eff}} \ge 120,000$\,K black body, a temperature necessary to explain the presence of the  \nev\ line in the nebular spectrum and comply with the SED's observed shape.  We now know that this is no longer necessary, so 
The data were de-redden according to \citet{1998ApJ...500..525S} with a canonical ratio of total-to-selective  absorption,  $E(B-V)=0.03$\,mag and $R_V=3.1$.  %Now, we can use our continuum fit to the stellar spectra from HST-STIS  and achieve a much better temperature determination  since  the interstellar extinction law  is well-known  for the HST wavelength range. 
We simultaneously fit two black bodies to the observed spectral energy distribution by introducing best guesses for T$_{\mathrm {cool}}$, T$_{\mathrm {hot}}$,  
$r_{\mathrm {cool}}/D$, and $r_{\mathrm {hot}}/D$, where $D$ is the distance to the object, and calculating their best fit values. Since there is a gap between extreme UV and X-ray wavelength ranges, and since the slope of the X-ray data is not strictly that of a black body, we obtain three distinct solutions with similar $\chi^2$ by  varying the input parameters. Possible solutions are presented in Table\,\ref{tab:sedsolutions}.

\begin{figure*}[p]
\includegraphics[width=16.0cm,bb = 0 5 380 380, clip=]{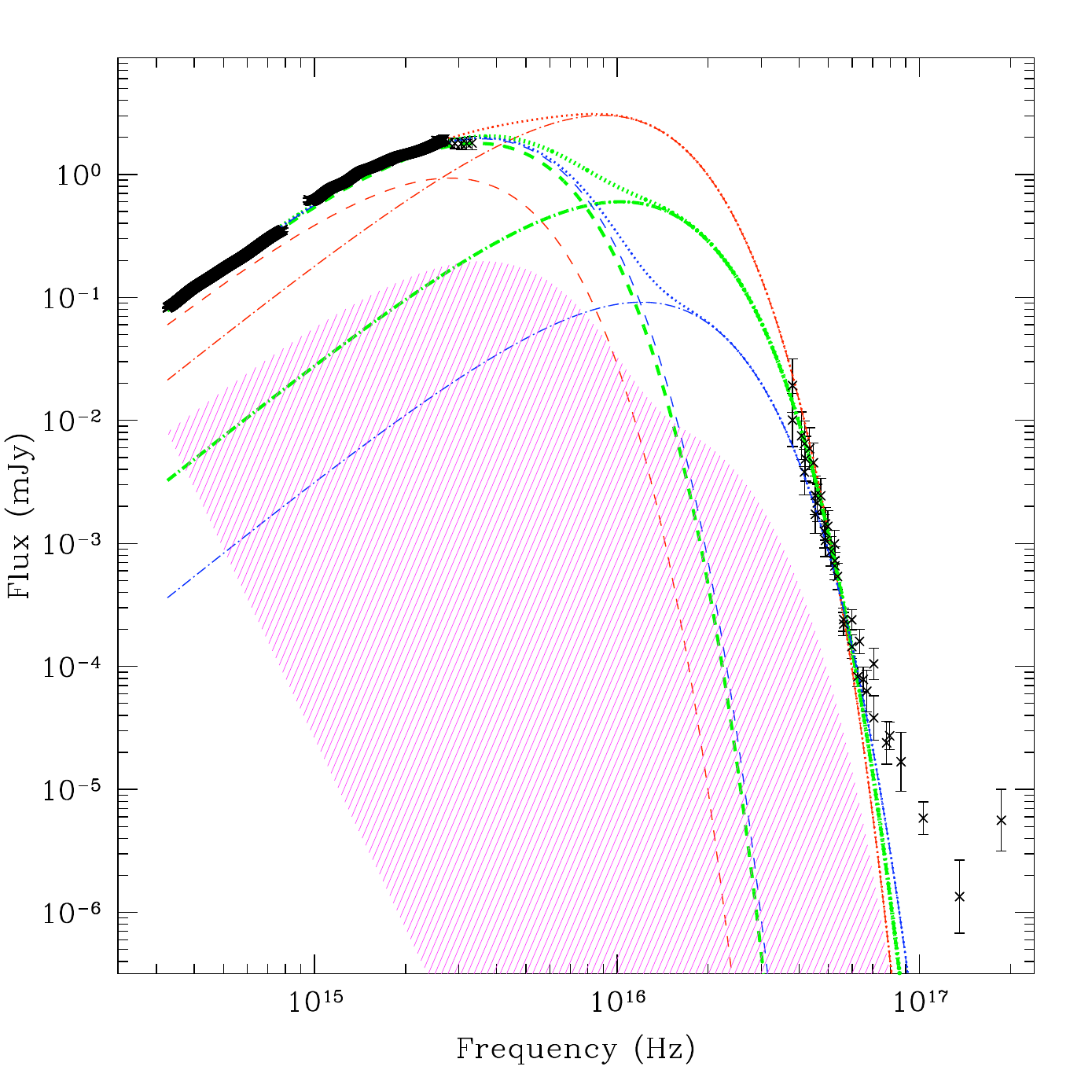}
\caption{The spectral energy distribution (SED) of \TS.  The observed spectra are presented by  crosses with error bars (SDSS + HST/STIS + FUSE+XMM).
The observations are fitted with two blackbodies. Three pairs of blackbody solutions are presented according to Table\,\ref{tab:sedsolutions}. Dashed, dashed-dotted and dotted lines represent the {\sl cool}, the {\sl hot} components and their sum respectively. 
%The  triangles and squares represent the fluxes of {\sl hot} and {\sl cool}  components respectively, as calculated for three possible configurations by {\sl Nightfall}. They are plotted to demonstrate that only a cool $\sim160\,000$ solution for {\sl hot} component satisfies simultaneous  SED, light curves  and RV fitting.
Shaded is the area in which  the the radius of  {\sl hot} component determined from the light curve analysis starts to deviate from the one required for SED fitting. In this area the  luminosity of  {\sl hot} component does not provide enough irradiation  to the  {\sl cool} component to produce observed light curve. Thus, only a cool $\sim160\,000$\,K solution is viable.
 }
\label{sed}
\end{figure*}

%------------------------------------------------------------------------------
\begin{table*}
\caption{Output parameters of double black-body  fits to the SED.}  
\label{tab:sedsolutions} 
%\scalebox{0.95}{
\begin{tabular}{lccc|ccccc}       
\hline 
\hline
   &  &  optical     component  & &   &  X    component  &  & & \\
 Solution  & T &  R/D  & D\tablenotemark{a}    &T  &   R/D   &  R\tablenotemark{a} & log  L\tablenotemark{a} &  $\chi^2$ \\
                  &  K       &  $\times10^{-13}$     &  kpc   &  K    &   $\times10^{-13}$   &   \rsun     &    \lsun              \\ 
    \hline 
  &&&&&&&&  \\
    Cool   &  47\,700 &  3.80   &  21.7    &  152\,600  & 1.20  &  0.12  &  3.81 & 2.54\\
    Intermediate & 58\,900 &  3.83    &  21.6    & 174\,600  & 0.43  & 0.04  & 3.14  & 2.63 \\
    Hot    &     60\,600       &   3.83  &  21.5    &  205\,200  & 0.15  &  0.014  & 2.51 & 2.65  \\
\hline   
\end{tabular}
\tablenotetext{a}{The values of these columns were calculated assuming R$_{\mathrm {cool}}=0.43$\rsun,  and corrected by a factor  $D \sim \sqrt{F}=0.85 $ because the model 
atmosphere flux $F_{\rm atm} \approx 0.73\times F_{\rm BB}$ with $T=T_{\rm eff}$ at the optical wavelengths, as it follows from the analysis below (see \ref{sec:cool} and \ref{sec:hot}).}
%\tablenotetext{b}{The luminosity of the hot component is calculated based on a distance from column 4 and assuming that both components emit as a black body. In reality the distance is overestimated. Using an atmospheric model for the cool component reduces the distance to $\sim$ 21 kpc, because the model 
%atmosphere flux $F_{\rm atm} \approx 0.73\times F_{\rm BB}$ with $T=T_{\rm eff}$ at the optical wavelengths, and $D \sim \sqrt{F}$. 
%The departure of the {\sl hot} component SED from black body is even greater in   
%the observed X-ray band and blackbody assumption leads 
%to overestimate of the luminosity. However, most estimates of temperatures and luminosities of super-soft X-ray sources are made using black bodies and these numbers are useful when comparing to other similar objects. }
%}
\end{table*}

The resulting fits are presented in Fig.\,\ref{sed}. %The {\sl cool} component for this new fit is much cooler than previously assumed with T$_{\mathrm {cool}} \approx60,000$\,K. 
The black body solution for the {\sl hot} component is T$_{\mathrm {hot}}\approx 180,000\pm25\,000$\,K.  Minimum  $\chi^2 \approx 2.6$ can be achieved with significantly different temperatures for the  {\sl hot} component, depending upon which part of X-ray data the  fit relies on.   %In both extremely cool as well as extremely hot  cases,  the Raleigh-Jeans tails of two black bodies have significantly different slopes, which helps in choosing between solutions.
But the hot solutions with T$_{\mathrm {hot}} \ge 175\,000$\,K, marked in the Fig.\,\ref{sed} as shaded area, do not work,  as can be seen below, because  only certain  ratios between the {\sl hot} and {\sl cool} components fluxes can produce the required difference in the depths of minima of the light curve. % If there is no irradiation of the hemisphere of the {\sl cool} component that faces its {\sl hot} companion, the dips in the light curve should be equal. The larger the temperature difference between components, the larger is the difference in the depths of the minima. The range of corresponding fluxes, the ratio of which can produce the observed light curve, are shown in the Fig.\,\ref{sed} as open squares and triangles for the {\sl cool}  and {\sl hot} components, respectively. They were calculated using the binary star modeling code {\sl Nightfall} \footnote{ http://www.hs.uni-hamburg.de/DE/Ins/Per/Wichmann/\ Nightfall.html.}. 
To restrict the range of possible solutions, we analyze the form of the light curve, together with the radial velocity curve, using the binary star modeling code {\sl Nightfall} \footnote{ http://www.hs.uni-hamburg.de/DE/Ins/Per/Wichmann/\ Nightfall.html.}.

{\sl Nightfall} is based on a physical model that takes into account the non-spherical shape of stars in close binary systems, as well as the mutual irradiation of both stars, and a number of additional physical effects such as gravitational darkening and albedo. We fitted simultaneously  the light curves in three bands and   the radial velocity curve.
The program uses differential magnitudes and is tailored to the Johnson photometric system. Taking into account that the shape and range of amplitudes of the light curve is (a) similar in the BUSCA narrow filters and the Johnson V filter and (b) does not show large wavelength dependence  in the optical  range, we assigned BUSCA {\sl uv} to the $U$ filter, {\sl b} to the $B$ filter, and {\sl r} to the $R$ filter. We did not use {\sl nir} band  data, since it  was the noisiest  and would not add anything substantial to the analysis.  We searched for solutions by setting the temperatures to the values estimated from the SED. We also fixed the total mass of the system close to the Chandrasekhar  limit of 1.39 \msun.  The real M$_{\mathrm{tot}}$ might be slightly  lower or higher, that would not affect this analysis. Leaving the total mass parameter free, {\sl Nightfall} tends to solutions involving massive stars, which are excluded.  However,   limiting the total mass still  results in a variety of remaining parameters that achieve similarly good fits. The inclination angle of orbital plane to the line of sight, $i$, must be relatively high to produce the observed light variation due to the ellipsoidal  form of the {\sl cool} component, but not too high to produce eclipses, which we do not observe, neither in the optical, nor in X-rays. Inclination angles ranging from the high 40's to the low 70 degrees  are acceptable, and the light curve form depends weakly on $i$ within that range. The essential parameters for which we seek solutions are the masses and radii of the  binary components. However, the RV and light curves fitting does not provide any clue on masses, so additional constrains were necessary.

\subsubsection {The {\sl cool} component.}
\label{sec:cool}

\begin{figure}[ht!]
\includegraphics[width=8.cm,bb = 60 60 800 550, clip=]{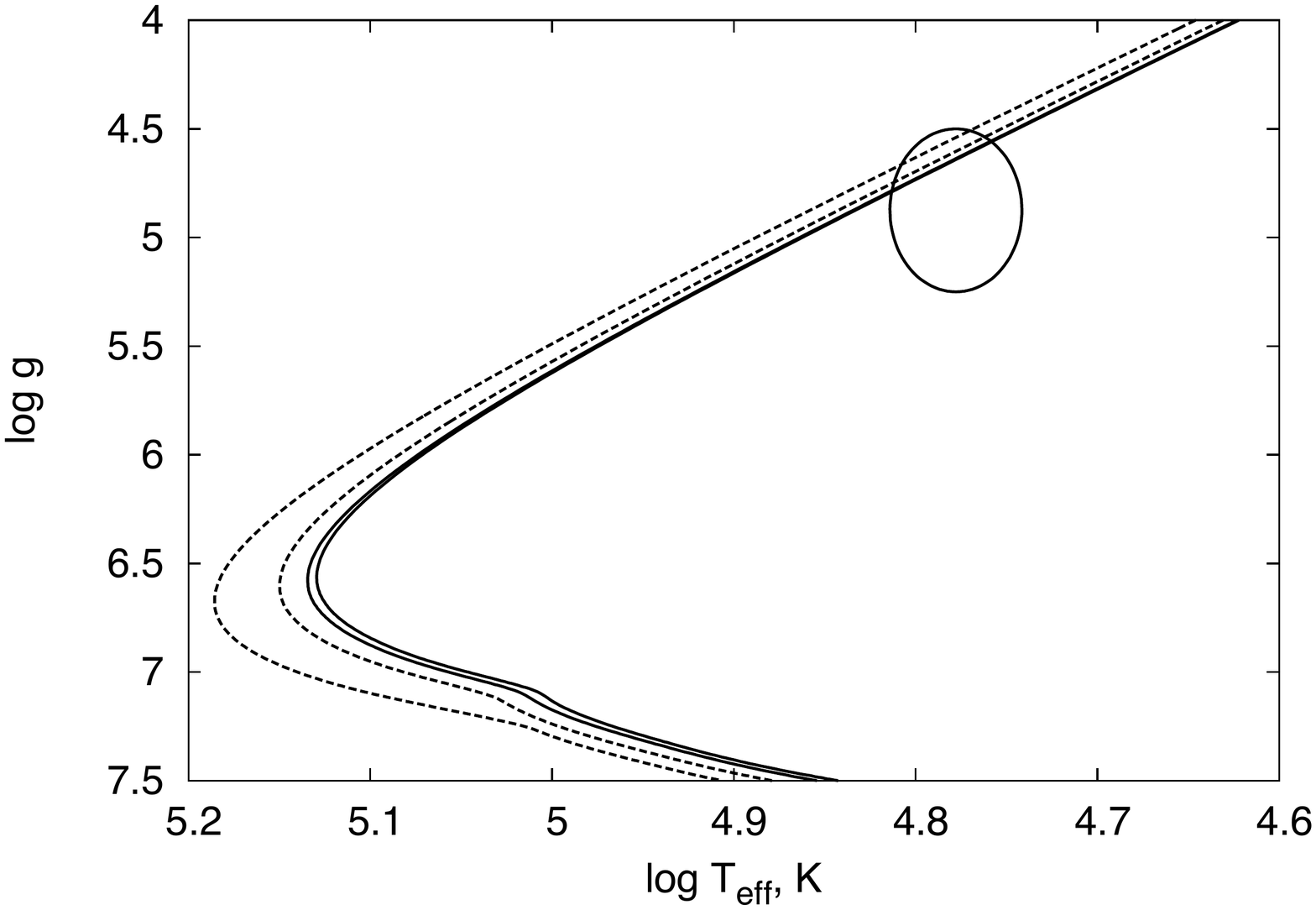}
\caption{ \te - $\log g$ dependence for post-AGB models  
\citep[][ and priv. comm.]{2009A&A...508.1343W}
Solid line - Z=0.004 for 0.529 \& 0.533 \msun.
Dashed line - Z=0.0005;  0.539, 0.551 \msun\ corespondingly. Tracks for more massive stars spread out toward hotter temperatures.  
The ellipse indicates range of possible solutions for \TS\ stemming from variety of methods used in this paper.}
\label{fig:te-g}
\end{figure}

%We have a lot more observational information on the {\sl cool, optical} component than on
%the {\sl X}-component. Its temperature is quite well determined, once we discovered that the nebular  \nev\ does not have to be caused by this star. The {\sl cool} component's %temperature from the continuum fit is $60\,000 \pm 5\,000$\,K.  A black body does a good job of determining the temperature of the {\sl cool} component,  since the effects of %atmospheric absorption due to chemical composition start playing a significant role 
%mostly beyond FUSE range. 

The first estimate of the mass of {\sl cool} component 
\citep{2008AIPC..968...62T,2007arXiv0709.4016T} based on 
$T_{\rm eff}$ and evolutionary tracks for solar composition post-AGB stars from
\citep{1983ApJ...272..708S,1995A&A...299..755B}
led to $M_{cool} \approx 0.58$\,\msun. Recent models
for different metallicities \citep{2009A&A...508.1343W} suggest
0.52\,\msun\  as the lower limit of the mass of a star that heats up to
$\sim 60\,000$\,K (see Fig.\,\ref{fig:te-g}). 
For Z=0.001, the mass limit is actually  0.54\,\msun.
%The first  attempts to estimate the  binary parameters \citep{2008AIPC..968...62T,2007arXiv0709.4016T}  led to an estimate of the mass of the {\sl cool} component, $M_{cool}$, of about  0.58\,\msun.  This was estimated using the post-AGB tracks of  \citet{1983ApJ...272..708S,1995A&A...299..755B}.  When applying the recent models that have been calculated by \citet{2009arXiv0903.2155W} for different metallicities,  the lower limit to the stellar mass that heats up to $\sim 60\,000$\,K can be as low as 0.52\,\msun  
%(see Fig.\,\ref{fig:te-g}). For a metallicity of Z=0.001, the mass limit is actually  0.54\,\msun. %\  and it is not substantially different from the mass estimate used previously or in \citet{paperI} considering the orbital dynamics or the ionization of the nebula, respectively.  
Fixing the mass of the {\sl cool} component  in {\sl Nightfall} to that value, we find that regardless of other poorly-constrained  parameters of the {\sl hot} component, the {\sl cool} star must have a radius of at least  0.42\,\rsun\  to fill its Roche lobe up to 94-99\% in order  to produce the observed light curve. Since the cool component is ellipsoidal in shape, this radius, as determined by {\sl Nightfall},  represents the mean radius. In fact, the radius  depends only weakly on the mass adopted for the star, and is a stronger function of the binary system's mass ratio, which determines the size of Roche lobe. For range of mass ratios stemming from the total mass of the system in 1.3--1.45\,\msun\ interval the mean radius of the {\sl cool} component lies within 0.42--0.45\,\rsun\ range.

The mass and radius obtained for the {\sl cool} component leads to a mean $\log g $ of $ 5.03\pm0.03$, a value deduced by averaging the unevenly distributed gravitational acceleration on its surface 
\citep{1992Ap&SS.196..241D}. A very similar value of  surface gravity is obtained by using FITSB2 (\S 3.1). 
%The model atmosphere spectra used for the    FITSB2  analysis were computed with the non-LTE code Pro 2 \citep{1986A&A...161..177W} and used only hydrogen and helium lines in the optical range.

%Given that the {\sl cool} component of TS01 nearly fills its Roche lobe and 
%that it is currently contracting, it has only recently terminated a 
%phase of common envelope evolution.  Its atmosphere, therefore is   
%clearly very different from that of a single star passing through the early   
%epochs  of its planetary nebula stage.  As a result, it might be expected to be 
%considerably smaller than a single star at the same evolutionary phase. 
%Indeed, the theoretical models of single stars with initial masses of 
%1.0\,\msun\  and metallicity Z=0.001 from \citet{1994ApJS...92..125V} 
%and \citet{2009arXiv0903.2155W}, shown in Fig.\,\ref{evol_sbs}, have larger radii and 
%smaller surface gravities than we observe.  This comparison clearly 
%indicates that a more complete analysis of TS01 is warranted than can be 
%made with single star models.  
%Below, in Section 4, we present a detailed evolutionary  scenario for  \TS\ and present more arguments concerning our mass estimates.

\begin{figure*}[t]
\includegraphics[width=7.5cm, bb = 62 550 250 755, angle=90, clip=]{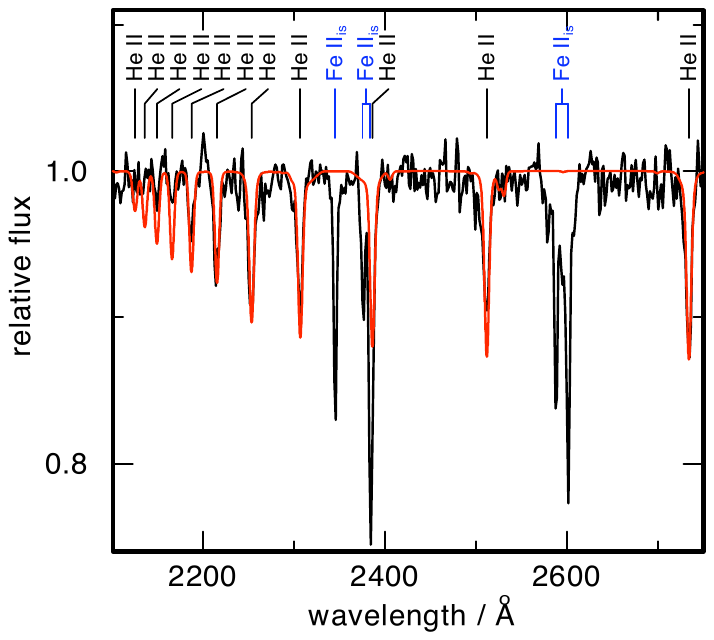}
\includegraphics[width=7.5cm,bb = 62 550 250 755, angle=90, clip=]{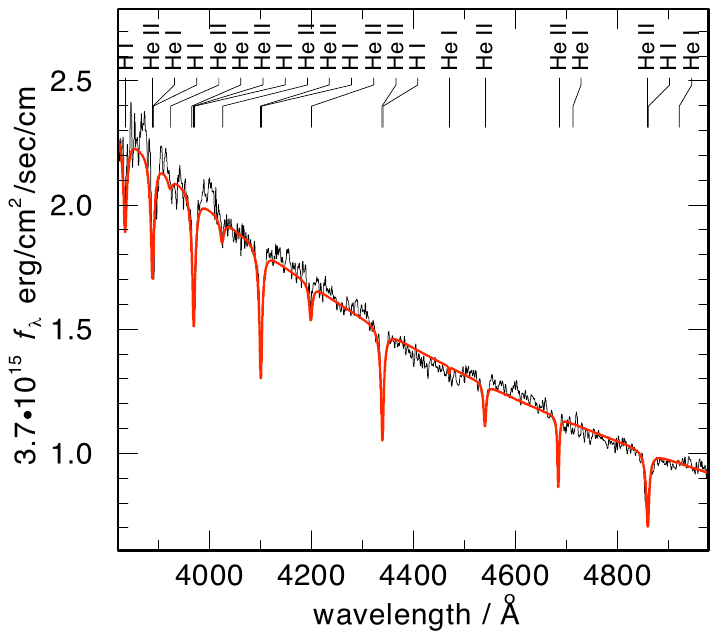}
\caption{ Comparison of a {\sl TMAP} SED with  observation.
a. the STIS NUV range. The He\,{\sc ii} Fowler series (marked at top) are well reproduced;
``is'' denotes interstellar lines.
b. the  GEMINI North optical range.
The H and He lines are marked at top.}
\label{trauch}
\end{figure*}

%Another way to estimate the stellar parameters is through a model atmosphere analysis.
Next,  we modeled the stellar atmosphere of the {\sl cool} component using the Tubingen NLTE Model-Atmosphere Package (TMAP)\footnote{
http://astro.uni-tuebingen.de/\raisebox{.3em}{{\tiny $\sim$}}rauch/TMAP/TMAP.html\hfill\hbox{}}  \citep{2003ASPC..288...31W,2003ASPC..288..103R}.  This code computes plane-parallel or spherical Non-LTE model atmospheres  in radiative  and hydrostatic equilibrium and considers opacities of all species from hydrogen to nickel. 
The determination of $T_\mathrm{cool}$ is based on the evaluation of the ionization 
equilibrium through the C\,{\sc iii}\,$\lambda$\,1175\,\AA\,/\,C\,{\sc iv}\,$\lambda$\,1169\,\AA\
stellar absorption line ratio. We find  $T_\mathrm{cool} = 55\pm5$\,kK.
At such a temperature, a surface gravity of  $\log g_{\mathrm {cool}} = 4.9\pm0.5$ gives a good agreement with the
observed spectral line profiles (see  Figs\@.\,\ref{trauch} a \& b, comparing a  {\sl TMAP} model with the \textsl{STIS NUV} observations  and with the \textsl{gemini} spectrum,
respectively).   $T_\mathrm{cool}$ and  $\log g_{\mathrm {cool}}$ cannot be better constrained given the quality of the data. However, the agreement between the values derived with  {\sl TMAP} and those derived with \textsl{FITSB2} and {\sl Nightfall} confirms 
our correct assessment of the basic parameters of the {\sl cool} component.

%We modeled the stellar atmosphere of the {\sl cool} component using the Tubingen NLTE Model-Atmosphere Package (TMAP)\footnote{
%http://astro.uni-tuebingen.de/\raisebox{.3em}{{\tiny $\sim$}}rauch/TMAP/TMAP.html\hfill\hbox{}}  \citep{2003ASPC..288...31W,2003ASPC..288..103R}, that is capable to calculate plane-parallel and spherical, chemically homogeneous, Non-LTE model atmospheres in radiative  and hydrostatic equilibrium and considers opacities of all species from hydrogen to nickel.
%For hot stars, the determination of $T_\mathrm{eff}$ is generally based on the evaluation of ionization 
%equilibria, i.e\@. spectral lines of successive ions have to be identified. 
%For \TS, this is only the case
%for C\,{\sc iii}\,/\,C\,{\sc iv}.  From C\,{\sc iii}\,$\lambda$\,1175\,\AA\,/\,C\,{\sc iv}\,$\lambda$\,1169\,\AA,
%we conclude  $T_\mathrm{eff} = 55\pm5$\,kK \citep{paperI}.
%At this $T_\mathrm{eff}$, a surface gravity of $\log g = 4.9\pm0.5$ gives a good agreement with the
%observation. In Figs\@.\,\ref{trauch} a \& b, we compare a SED
%calculated from a {\sl TMAP} model atmosphere with the
%He\,{\sc ii} Fowler series in the NUV and with our optical spectrum obtained with Gemini-North,
%respectively.  $T_\mathrm{eff}$ and $\log g$ cannot be better constrained because there are 
%not many lines and the S/N of the data is not sufficient to detect metal lines in this
%extremely metal poor star. However, the agreement of the {\sl TMAP} SED with the observation confirms 
%our correct assessment of basic parameters of the {\sl cool} component of the binary.

 \begin{figure*}[p]
\includegraphics[width=14.0cm,bb = 0 5 380 380, clip=]{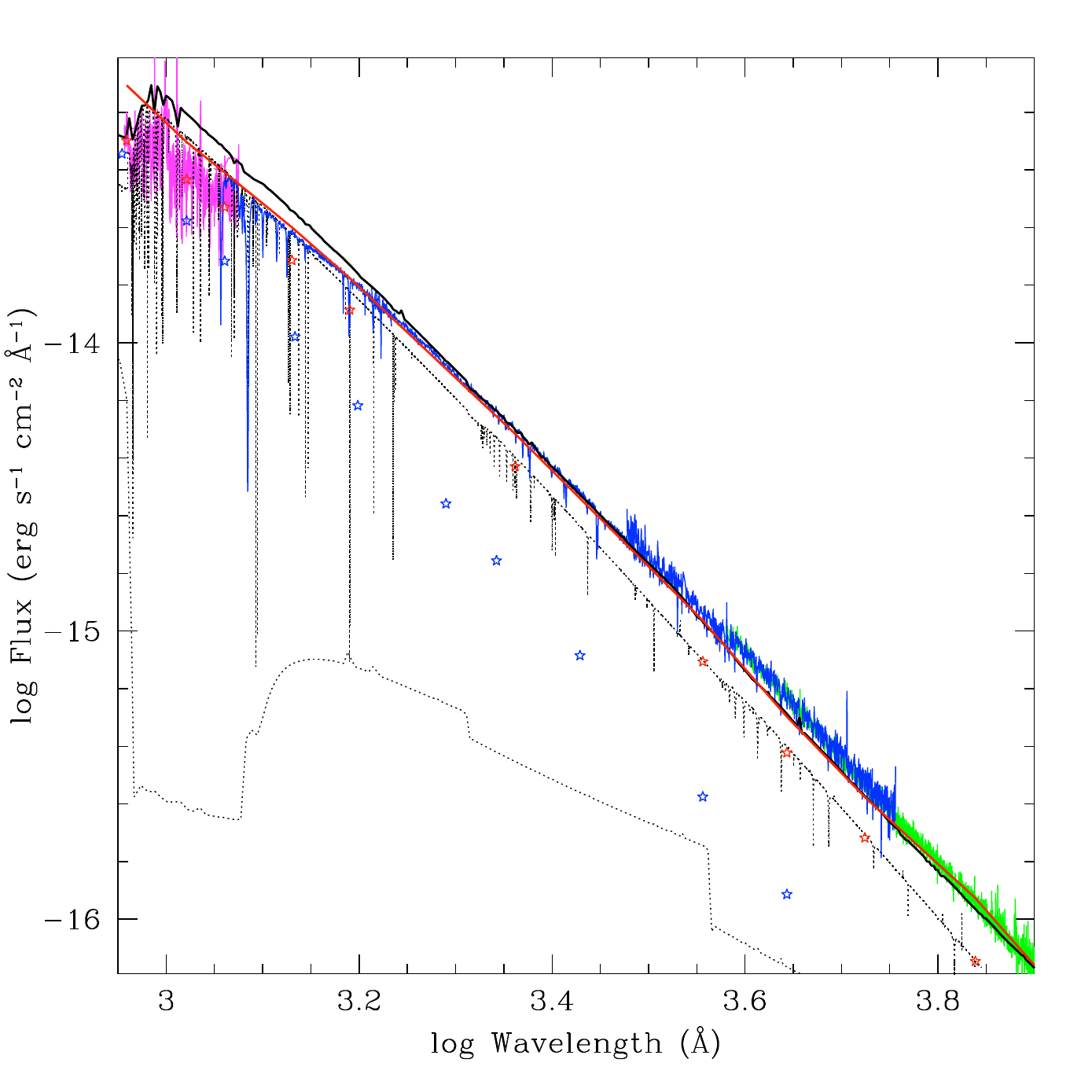}
\caption{The spectrum of \TS\ from the far UV to the near IR. The observational data after extinction correction and nebular emission lines removal 
%using $R=3.1$ and $E(B-V)=0.03$\,mag, 
are denoted as follows: FUSE - magenta; HST (FUV +NUV+UVopt)  - blue;  SDSS - green. The models for the nebular and stellar  emission are shown  by grey dotted lines.
%: the stellar atmosphere is from the present work while the nebular model is from \citet{paperI}. 
The thick black line is the sum of the model continuum emission from the nebular and stellar components. The black-body solutions are presented by red open stars for the {\sl cool} component and by blue stars for the  {\sl hot} component.  The thick red line is sum of {\sl hot} (162\,kK) and {\sl cool} ($\sim 57$\,kK) blackbodies plus nebular emission.}
\label{composite_spectra}
\end{figure*}

\begin{table}
\caption{Photospheric abundances of the {\sl cool} component of \TS.
         Ca-Ni are represented by a generic model atom.
         The errors for H, He, and C are about 0.3 dex.
         For N, O, Si, and the iron-group elements upper limits are given.}  
\label{tab:abundances} 
\begin{tabular}{cr}  
\\     
\hline
element & mass fraction  \\ 
\hline
H       &       7.471E-01 \\
He      &       2.525E-01 \\
C       &       1.335E-04 \\
N       & $<$\, 8.306E-05 \\
O       & $<$\, 7.116E-05 \\
Si      & $<$\, 6.737E-05 \\
Ca-Ni   & $<$\, 1.319E-06 \\
\hline   
\end{tabular}
\end{table}

We performed some  {\sl TMAP} test calculations in order to derive upper abundance limits for some
metals. These are summarized in Tab\@.\,\ref{tab:abundances}. 
Note that the resonance lines of C\,{\sc iv} and N\,{\sc v} were not used in our abundance determination, 
since they were found to be affected by interstellar line absorption. The chemical composition 
of the {\sl cool} stellar component is close to that of the nebula \citep{paperI}.

 \begin{figure*}[htp]
  \begin{center}
 \includegraphics[width=5.4cm,bb = 15 10 380 390, clip=]{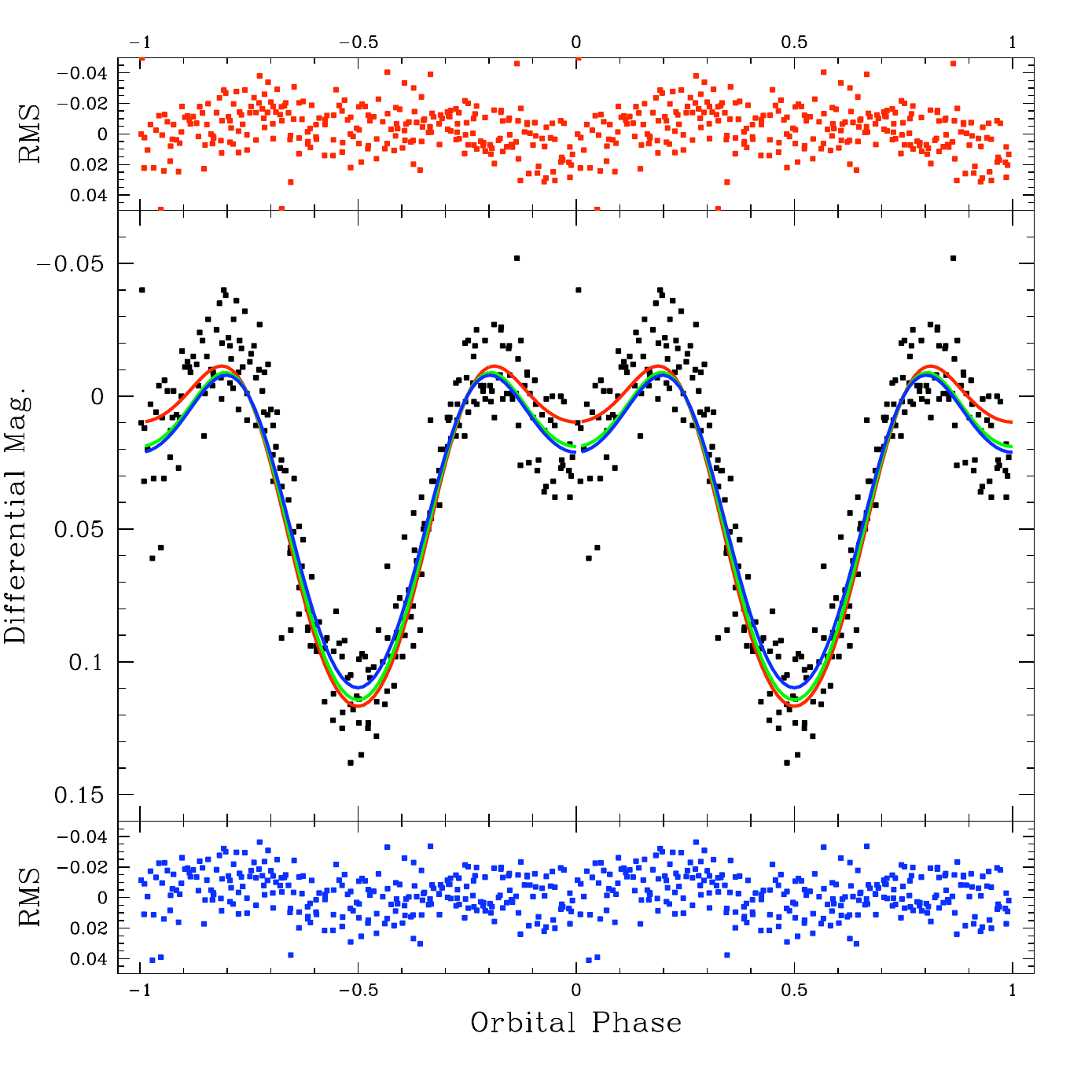}
 \includegraphics[width=5.4cm,bb = 15 10 380 390, clip=]{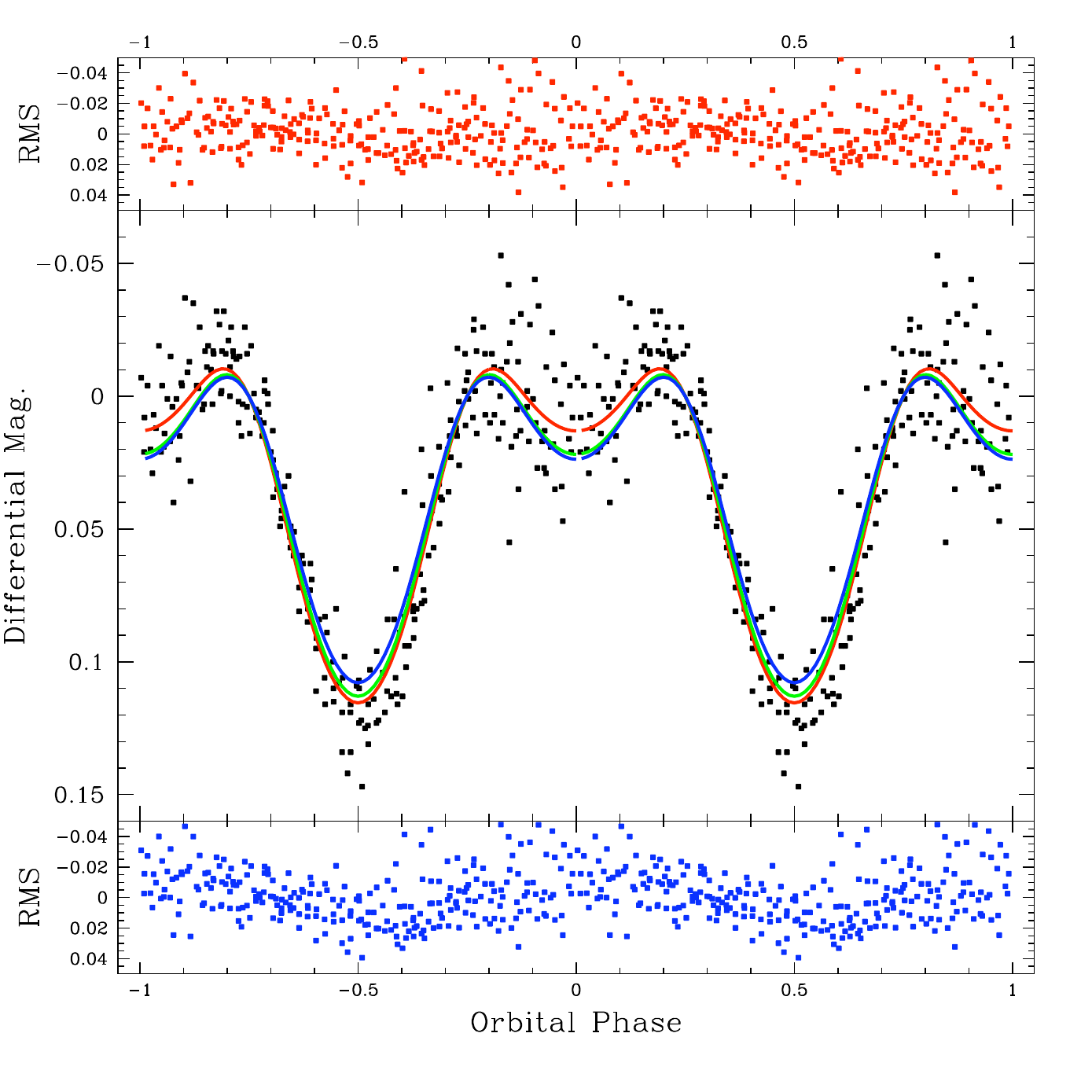}
 \includegraphics[width=5.4cm,bb = 15 10 380 390, clip=]{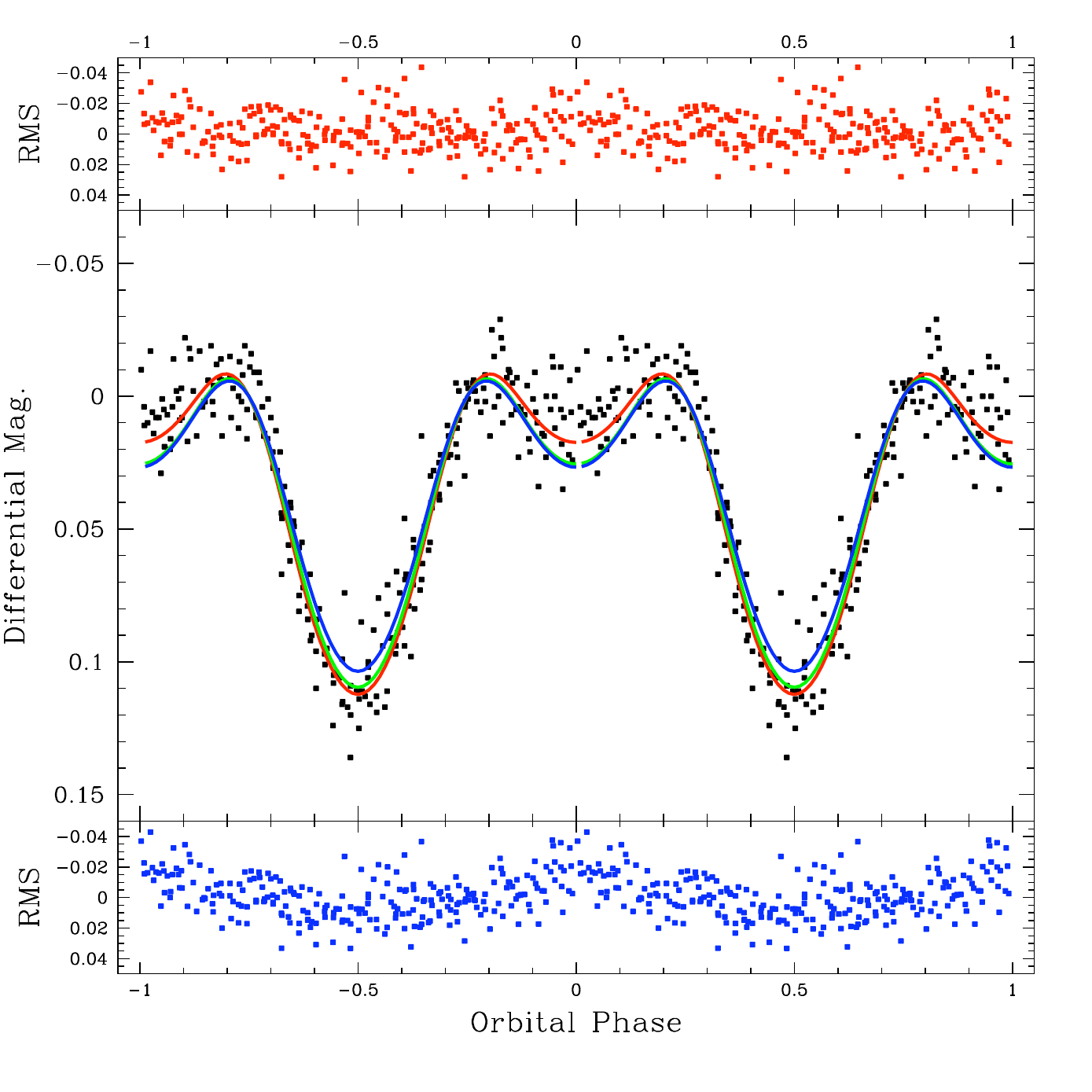}
 \end{center}
\caption{From left to right, the {\sl Nightfall} fits to the light curves in {\sl uv, b}, and {\sl r}. In the main panel, the red, green, and blue curves represent the solution for X-component with T$_{\rm eff}$ of 162\,kK, 182\,kK, and 202\,kK, respectively.  The deviations with respect to the coolest and hottest solutions are shown in top and bottom panels, respectively. The complete set of parameters used in fitting is given in Table\,\ref{tab:NFparameters}. The simultaneous fit to the RV curve is similar to the {\sl sin} curve presented in Fig.\,\ref{rv}.}
\label{nightfall}
\end{figure*}

In  Fig.\,\ref{composite_spectra}, the model is compared with observations, covering continuously the whole range from 900 to 10\,000 \AA.  Apart from the observations and atmospheric model, the nebular emission is shown as deduced in \citet{paperI}, and the black body curves as implemented in {\sl Nightfall}.  The black body corresponding to the {\sl cool} and  {\sl hot} components are denoted by open stars (red and blue respectively). 
The sum of the two black bodies and the nebular continuum  within the observing slit as computed in  \cite{paperI} is represented with the thick red line in Fig. \ref{composite_spectra}.  The fit of the models to the observations is excellent from the near infrared  to 1500 \AA.

The  curve representing the stellar model plus the nebular emission departs slightly from the observations at shorter wavelengths. The problem with the flux  above 1500 \AA\ in the NLTE models has been noted before \citep{2008A&A...481..807R}. More importantly, the model is calculated for a spherically symmetric star with a homogeneous temperature distribution over its surface. 
However, we know that the {\sl cool} component is strongly  irradiated and gravitationally distorted, which affect both its SED and the gravitational acceleration over the surface of the star. We approximate the observed, phase-averaged spectrum by the non-irradiated atmosphere model even though, in some orbital phases, we observe the irradiated hemisphere of the {\sl cool} component.  The spectra of   
irradiated atmospheres are flatter in the far UV in comparison with the spectra of non-irradiated atmospheres with the same $T_{\rm eff}$ (see next subsection). Therefore, the average flux in the far UV decreases in comparison of the non-irradiated atmosphere spectrum.  Overall, we have very good agreement between models and observations. 

%------------------------------------------------------------------------------
\begin{table*}
\caption{Parameters of {\sl Nightfall}  fits.}  
\label{tab:NFparameters} 
\scalebox{0.95}{
\begin{tabular}{lcccc|c|cccccccc}       
\hline 
   &  &  opt/UV & component     &  &  &   &  X &  component   & &  &  & \\
 Solution  & T &  M  &   R  & log $g$    &  inc & T &  M  &   R  & log $g$ &    $\chi^2$ \\
                  &  K  &   \msun     &  \rsun      &                &     deg                &  K    &   \msun   &   \rsun     &                &            \\ 
    \hline 
    Cool   &  57\,100  &   0.537  & 0.44 &  5.02 &   52.8  & 162\,195 & 0.853 &0.135 & 6.1 &  46.27 \\
    Intermediate & 57\,264 & 0.537 & 0.43 & 5.04& 53.0 & 182\,195 & 0.853 & 0.113 & 6.3 & 46.37 \\
    Hot    &              57\,104 & 0.537 & 0.43 & 5.04 & 53.2 & 202\,000 &  0.853 & 0.088 & 6.5 & 46.43 \\
\hline
\end{tabular}
}
\end{table*}

\subsubsection {The hot component.}
\label{sec:hot}

The parameters of the {\sl hot} component are less certain. Our knowledge  of the  {\sl hot} component is based on the binary period, the fact  that the {\sl cool} component is partially irradiated to produce the observed light curve, and the X-ray flux, which cannot originate from  the {\sl cool} component.  There is an extensive argument in \citet{2005AIPC..804..173N} discussing why the  {\sl hot}  component should be a compact object with a mass exceeding the mass of the {\sl cool} component. However, neither the light curve, nor the X-ray spectra allow us to determine the  temperature or  radius of the {\sl hot} component as well as we did for the {\sl cool} component.  

%Stellar atmosphere models depart significantly from the black body approximation at higher energies. The difference depends greatly on the chemical composition. Pure hydrogen or H/He models produce huge excesses at high energies, which are not observed in X-ray spectra. Adding metals brings the flux down, but the slope of  the models, regardless  of the chemical composition, never reaches good agreement with  observed spectra.  The assumption that the chemical composition of the envelope of accreted matter around the {\sl hot} component is similar to the nebular one and/or to that of the donor {\sl cool} component is not necessarily valid, because nuclear burning on the surface of the {\sl hot} component should have produced significant changes, as are often observed in novas.

We face two problems in the case of \TS's {\sl hot component}.  First, \TS\ is only detected in short, soft end of the X-ray range. Second,  the calibration of data at the extreme soft end of the XMM-{\sl Newton} spectral range is not very reliable.  Both prevent us from fitting exact atmospheric models to it.  Hence, like most studies of supersoft X-ray sources, we are forced to continue the analysis using the blackbody that successfully describes the spectrum of the {\sl hot}  component in the optical and UV range up to $4\times10^{15}$\,Hz (see Fig.\,\ref{sed}). 
We also note that  this introduces an
overestimate of the luminosity of the X-ray source
\citep{1994A&A...288L..45H, 2002ApJ...574..382S}. On the other hand
the temperature can be either overestimated
\citep{2002ApJ...574..382S}, or underestimated \citep{1994A&A...288L..45H,2003ARep...47..186I}.
Therefore, when estimating the temperature of the {\sl hot} component in the optical/UV range separately from estimates in X-rays, we should not worry  too  much if discrepancies arise.    It is  within the optical/UV range  that the irradiation of the {\sl cool} component matters, so we may fix the parameters of the {\sl cool}  component in {\sl Nightfall} and seek solutions for the {\sl hot} component.  Even so, varying freely both the radius and temperature  of {\sl hot}  component, we still do not reach unambiguous solutions.

In Fig.\,\ref{nightfall}, we present the fits produced by {\sl Nightfall} to  light curves in three different filters.  The parameters for the fits for the different temperatures of the {\sl hot} component  are presented in Table\,\ref{tab:NFparameters}. The temperature and the mass of the {\sl cool} component were kept fixed (the difference in mean temperatures of  the  {\sl cool} component in the Table reflects the different degree of irradiation). The parameters that were fitted are the orbital inclination angle, the fraction to which both components fill their Roche lobes, and the temperature  of the {\sl hot} component.  Three models with different temperatures for the {\sl hot} component are  displayed in  Fig.\,\ref{nightfall}.  The fits shown red, green, and blue correspond to T$_{\rm hot}$= 162, 182, and 202\,kK, respectively.  The lower and upper panels show the deviations  of the fit from the observations for the extreme cold and hot solutions.  %These are the same solutions for  the  {\sl hot} component that are shown in Fig.\,\ref{composite_spectra} by dotted lines, with similar color coding as in Fig.\,\ref{nightfall}. These are also the same  {\sl Nightfall} solutions presented in Fig.\,\ref{sed}. The hotter the source of irradiation, the fainter it should be in the optical range, resulting also in a reduced  size of the {\sl hot} component. 
As can be seen from Fig.\,\ref{nightfall} and Table\,\ref{tab:NFparameters} the differences between three models ranging from 160 to 200\,kK are not important in the optical domain. 
%But  cross-checking results obtained from SED fitting and {\sl Nightfall} modeling  confine the cool T$_{\rm hot} \approx160\,000$\,K solution, as the only one satisfying both methods. 
 
 Note, that the range of temperatures for {\sl hot} components obtained from blackbody fitting to SED and from the fitting of light curves by {\sl Nightfall} are  similar.  But comparison of Tables \ref{tab:sedsolutions}  and \ref{tab:NFparameters} shows that only for low temperature solution the radii deduced by both methods are compatible. 
Introducing the radius of the {\sl cool} component deduced from {\sl Nightfall} into the $R_{\mathrm cool}/D$ parameter used in the fit of two black bodies to the SED %presented in  Fig.\,\ref{sed} 
results in a distance of 25 kpc for the minimum radius of R$_{\mathrm cool} = 0.42$ \rsun\  and leads to $R_{\rm hot} \approx 0.13 R_{\odot}$ for a   blackbody temperature 150\,kK. % The solutions with larger temperatures from  the Table\,\ref{tab:sedsolutions} yield disparities between the luminosity of   the {\sl hot} component deduced from the SED fitting and from  {\it Nightfall}, as shown in Table\,\ref{tab:NFparameters}. 
The color temperature of   the {\sl hot} component  is probably slightly higher. This is  caused by the divergence of the real atmosphere   from the black-body at high energies and also  by the use of a Roche lobe-shaped {\sl cool} component in {\sl Nightfall}, instead of the spherical shape in all other calculations.  Similarly, using stellar atmospheres instead of black-bodies reduces the  distance to about 21 kpc, as reflected in  Table\,\ref{tab:sedsolutions}. Increasing temperature is compensated naturally by a smaller radius.
The solutions with temperatures above 175\,kK come up with parameter R$_{\mathrm{hot}}$/D  too small to be compatible with results of light curve fitting.  In Fig.\,\ref{sed} the blackbody solutions of  {\sl hot} component that fall into shaded area are not luminous enough to provide necessary irradiation and produce the observed light curves.  
In the meantime, temperatures above 185\,kK are not tolerated by ionization modeling of the nebula \citep{paperI}.

%\begin{figure}[ht]
%\includegraphics[width=11.cm,bb = 50 100 1050 500, clip=]{tovmassian_fig14}
%\caption{$\chi^2$ map of {\sl Nighfall} T$_{\mathrm {eff}}$/R$_{hot}$ solutions. The dark parabolic-shaped strip indicates the locus of the lowest $\chi^2$ values. The vertical line correspond to the upper limit for temperature of the {\sl hot} component, constrained from the X-ray observations.} 
%\label{chi2nightfall}
%\end{figure}

%The fit to the radial velocity curve by  {\sl Nightfall} is very similar to the one presented in Fig.\,\ref{rv}.  In fact, solutions for low and high total mass are indistinguishable in such a plot. Overall, the $\chi^2$ value does not change much from one set of parameters  to the other. The set of solutions with similarly small $\chi ^2$ is presented in Fig.\,\ref{chi2nightfall}.  The dark strip indicates the lowest  $\chi ^2$ solutions and corresponds to a luminosity of $\approx 10^4   L_{\odot}$.   This luminosity is necessary to explain the observed optical 
%light curve (i.e., the degree of irradiation).  We can effectively impose an upper limit of 250 kK, because, with higher temperatures, the {\sl hot} component would have detectable X-ray flux at higher frequencies.  The low temperature cut is at 160\,kK, but the best solutions are actually close to 160\,kK.  The coolest solution obtained  from the SED fitting (see Table \ref{tab:sedsolutions}) is the only one that yields a similar luminosity for the {\sl hot} component.   
Therefore, we consider  160--175\,kK temperature range and R$_{\mathrm{hot}}\approx0.1$\rsun\ to be the closest to the real   
properties  of the {\sl hot} component.  Note that the {\sl hot} component has a radius R$_{\mathrm{hot}}\ge 0.04$\rsun\, at least. This is much larger than an ordinary white dwarf.    As such \TS's {\sl hot} component is very similar to the supersoft X-ray source (SSS)  Lin 358, one of the two SMC symbiotic stars studied by \citet{2007ApJ...661.1105O}.  %The combination of temperature and radius  corresponding to the cool solution is the only one providing luminosity $\approx 10^4   L_{\odot}$ for the hot component, necessary to produce observed irradiation.
Majority of estimates of temperatures and luminosities of super-soft X-ray sources are made using black bodies and values obtained here  are useful when comparing to other similar objects.

 \begin{figure}[ht!]
\includegraphics[width=7.5cm,bb = 1 100 550 790, clip=]{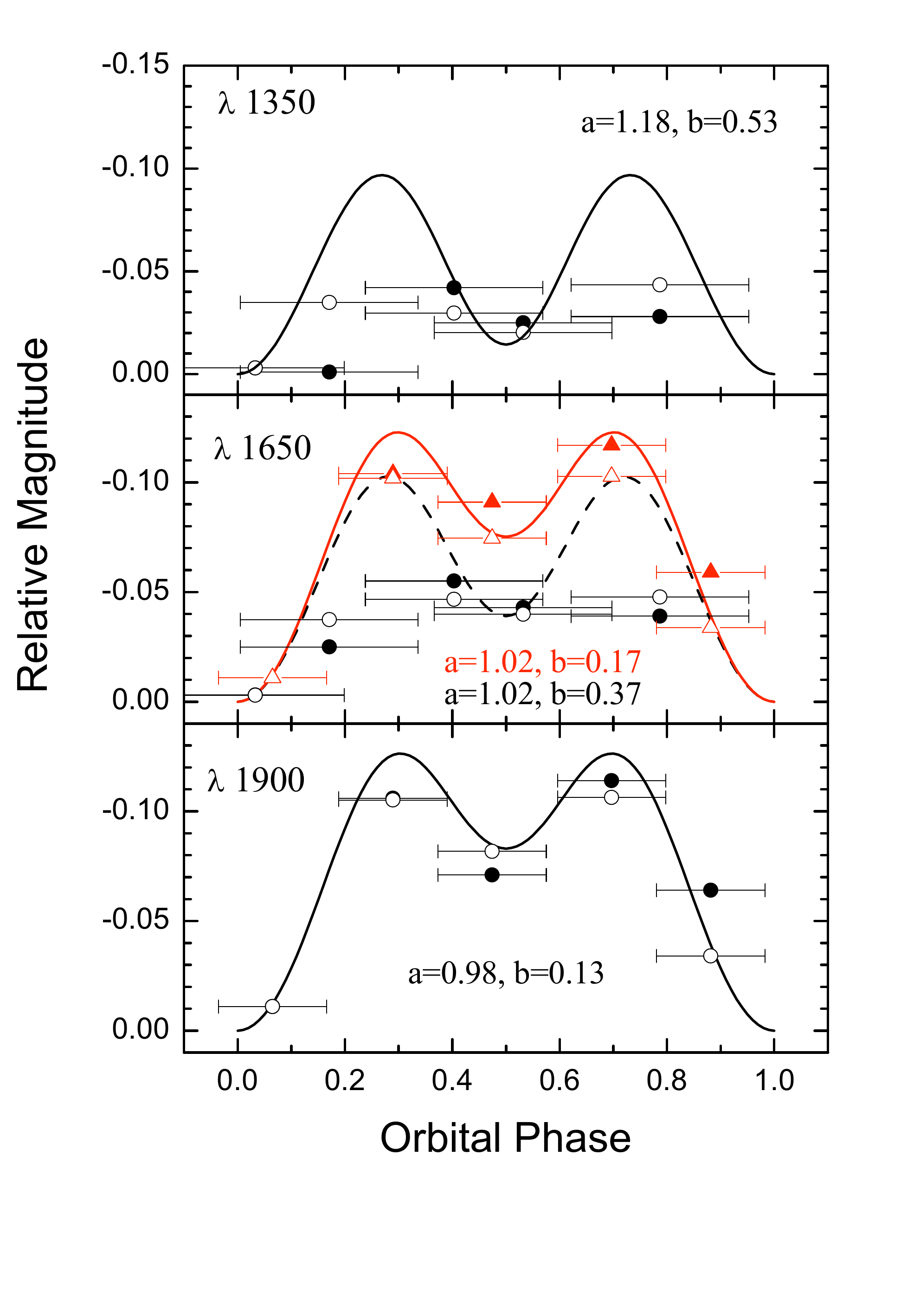}
\caption{ Comparison of the observed light curves (filled symbols) at $\lambda$1350\,\AA\ (top panel), 
$\lambda$1650\,\AA\ (middle panel), and $\lambda$1900\,\AA\ (bottom panel) with
the computed light curves in the same bands (solid and dashed lines). The horizontal error bars mark exposure length for each point. 
The  light curves at $\lambda$1650\,\AA\ are obtained separately from  {\it HST} FUV (circles) and NUV (triangles) detectors. 
Their disparity  indicates the difference of light curve smoothing due to the exposure length. The model fluxes, integrated
in the same phase ranges as observed ones are shown by open symbols. All models are computed
for the {\it cool} solution from Table 2. To account for  differences between the black body and real atmosphere  we introduced the  
parameters $a$ \& $b$ (see text), obtained empirically from comparison of these two observed light curves.The  values of parameters  for flux correction  
in corresponding bands are marked in the plot.
} 
\label{uvlc}
\end{figure}

Fitting an exact atmospheric model to the X-ray data for  the {\sl hot} component does not make much sense, because the observed energy range is too small and the quality of the data is too poor.\footnote{For the analysis of  the ionization of the nebula, however, using a black body would cause  serious problems  at high energies. This is why, for their photoionization modelling, \cite{paperI} selected a suitable spectrum from a grid of models with halo composition, which reasonably well describes the observed fluxes both in opt/UV and X-ray range and  provides a more realistic  picture.}

An additional test for checking  the estimated temperature of the {\sl hot} component  comes from modeling of the light curve in the UV.  However, {\it Nightfall} cannot calculate model fluxes in the UV, so we implemented our own code \citep{2002ARep...46..656S}, which calculates the irradiation in a binary according to the prescription in \citet{1983MNRAS.202..347H}. {\it Nightfall} is based on the same algorithm, so we naturally obtained exactly the same light curves for optical bands.  %Using the cool  162\,kK solution from Table \ref{tab:NFparameters}, w
We compute the UV light curves for the cool soloution from Table \ref{tab:NFparameters}, and present it, together with the observed one,  in Fig.\,\ref{uvlc}.  In computing the  UV light curves we take into  account  the difference between a black body  and a stellar   atmosphere introducing the ratio $f_\lambda = F_\lambda ({\rm BB})/F_\lambda ({\rm BSA})$ of black body and a stellar atmosphere fluxes at a given wavelength\footnote{In the optical part of the spectrum  $f_{\rm opt} \approx$ 0.73 for both the \textsl{cool} and \textsl{hot} components; near the Lyman edge  the SED of the {\sl cool} component is  different from a blackbody: $f_{\rm 1350} \approx 1.18$ and $f_{\rm 1650} \approx 1.02$; the values of $f_{\lambda}$ for the {\sl hot} component remain close to 0.73. Moreover, the values of $f_{\lambda}$ for the {\sl cool} component must depend on the  irradiation flux. If the irradiation   flux increases, $f_{\lambda}$  decreases, because the spectra of the irradiated stellar atmosphere  are getting  closer to a black body spectrum.   The best approximation of the observed light curves by the model light curves was obtained for a simple linear dependence $f_{\lambda} = a - b \cdot (F_{\rm irr}/F_0)$, where $F_{\rm irr}$   is the irradiation flux at a given point of the {\sl cool} component surface, and $F_0$ is the flux from the {\sl cool} component at the same   point. A change of the continuum slope in the UV band at the various orbital phases, observed by HST, confirms this picture. Parameters $a$ and $b$ deduced for each $\lambda$ are presented in Fig.\,\ref{uvlc}.}.

%. The observed  and calculated UV light curves are presented in Fig.\,\ref{uvlc}.  In computing the  UV light curves we take into  account  the difference between a black body  and a stellar   atmosphere.  We introduce the relation between a black body and a stellar atmosphere fluxes as a $f = F_{\rm SA}/F_{\rm BB}$, which can have   different values for  different wavelengths for stars of different temperature: in the Rayleigh-Jeans/optical part of the spectrum of both stars $f_{\rm opt} \approx$ 0.73; near the Lyman edge  the SED of the {\sl cool} component is  different from a blackbody and $f_{\rm 1350} \approx 1.18$ and $f_{\rm 1650} \approx 1.02$; the values of $f_{\lambda}$ for the {\sl hot} component remain close to 0.73. Moreover, the values of $f_{\lambda}$ for the {\sl cool} component must depend on the  irradiation flux. If the irradiation   flux increases, $f_{\lambda}$  decreases, because the spectra of the irradiated stellar atmosphere  are getting  closer to a black body spectrum.   The best approximation of the observed light curves by the model light curves was obtained for a simple linear dependence $f_{\lambda} = a - b \cdot (F_{\rm irr}/F_0)$, where $F_{\rm irr}$   is the irradiation flux at a given point of the {\sl cool} component surface, and $F_0$ is the flux from the {\sl cool} component at the same   point. A change of the continuum slope in the UV band at the various orbital phases, observed by HST, confirms this picture. Parameters $a$ and $b$ deduced for each $\lambda$ are presented in Fig.\,\ref{uvlc}.

\subsection {The total mass.}

\begin{figure}[ht]
\includegraphics[width=8.1cm,bb = 10 0 750 590, clip=]{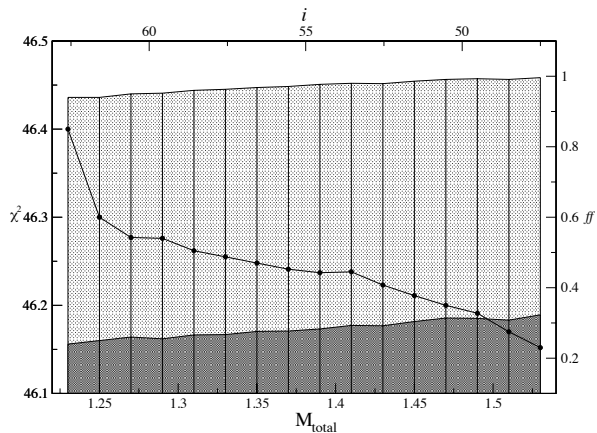}
\caption{The line with dots depicts  the  dependence of the $\chi^2$ value on the total mass of the system.  Fixed parameters in the calculations were the temperatures of the components (55 \& 160 kK respectively) and  M$_{\mathrm cool}$ (0.54 \msun).  The free parameters were  the  inclination of the binary orbit $i$, which changed similar  to the $\chi^2$ in a range of values shown on the upper axes; and the Roche lobe filling factors ($ff$) of both components shown as shaded areas in a range of values marked on the right side axes. The dark shaded area corresponds to Roche lobe filling factor of the {\sl hot} component and the light grey to that of the {\sl cool component}.  }
\label{mtot_chi2nightfall}
\end{figure}

The mass of the {\sl hot} component cannot be determined directly from  the observed data.  However, simultaneously fitting the RV curve and light curves  indicates a tendency towards  improvement as the total mass approaches  the Chandrasekhar limit. In Fig.\,\ref{mtot_chi2nightfall}, we plot the calculated  $\chi^2$ as a function of the total mass of the system, by fixing in {\sl Nightfall}  the temperatures of the components to  our best estimates, i.e. 57 and 162 kK and adopting a mass of 0.54\msun\  for the {\sl cool} component. 
The lower limit for  total mass is of 1.25\,\msun, corresponding to a {\sl hot} component with a mass of $\sim 0.7$\,\msun.  This mass  corresponds to the lower limit for a white dwarf to sustain steady nuclear burning on its surface.  Increasing the total mass of the binary from that minimum value produces a decrease in the $\chi^2$ value, with a small plateau of  $\chi^2$ at about 1.4\msun. %almost exactly at the Chandrasekhar limit. 
The $\chi^2$ keeps falling as the  total mass increases. This is due to the fact that rising mass of the {\sl hot} component shrinks the Roche lobe of the {\sl cool} star. At around M$_{\mathrm {total}}=1.47$ the {\sl cool} component fills its corresponding Roche lobe to $99.9\%$, which helps a better fitting of the light curves.  Improvement   of  $\chi^2$ from there on is conditioned by  the   rapid increase of  size (Roche lobe filling factor = $ff$)   of the {\sl hot} component, which might be unrealistic. The mass   could in principle be  constrained by the fit to the RV curve, but unfortunately the RV data is too poor in quantity and quality to have strong influence on  $\chi^2$.  We consider  the flattening of   the   $\chi^2 -  {\mathrm{ M_{total}}}$ curve around M$_{\mathrm {total}}=1.40$ as  an  indication of the best  solution, where a balance is achieved between fitting the light and  RV curves at the same time, but of course a lower than Chandrasekhar limit mass is not excluded.
Better  measurements of radial velocities are required for  a more  reliable determination of stellar masses in  \TS.

 \begin{figure}[ht]
\includegraphics[width=7.5cm,bb = 20 5 550 430, clip=]{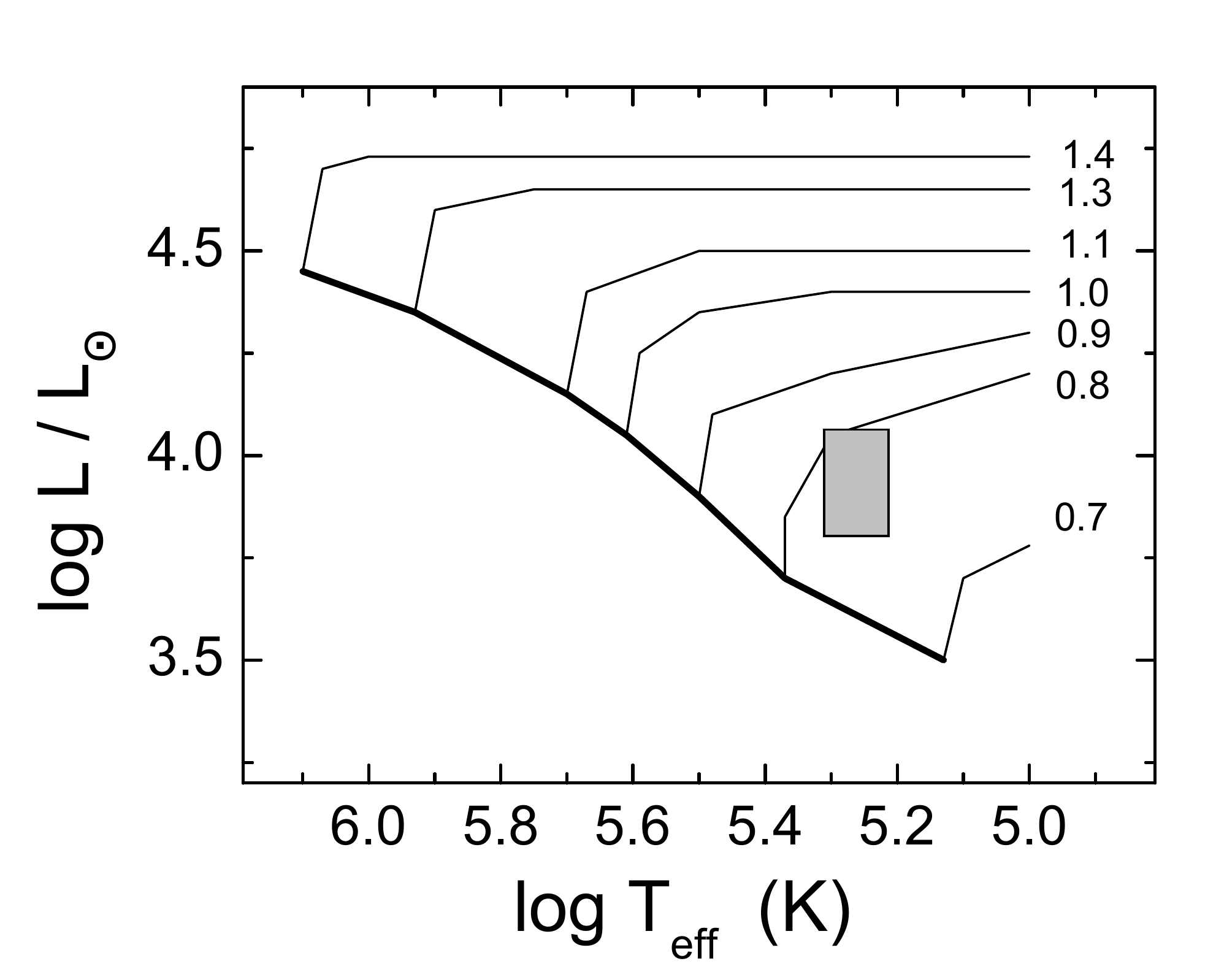}
\caption{ Position of the {\sl hot} component in the   
temperature-luminosity diagram (filled box) respective to the tracks   
of hot accreting white dwarfs in the 
steady-burning approximation 
\citep{1982ApJ...259..244I}.  The bold solid curve shows the   
high-temperature boundary of the stable-burning strip.  The numbers by   
the tracks are corresponding white dwarf masses in \msun. }
\label{hotHR}
\end{figure}

The mass estimate of the {\sl hot} component  can be checked from  its position in the H-R diagram  \citep{2003ARep...47..197S}. Figure~\ref{hotHR} shows position of the {\sl hot} component of TS01   
in the $\log$ T$_{\rm eff} - \log$\,L diagram. It suggests that the   
object has a mass of about $0.78 \pm 0.02$\,\msun, if it has not cooled much since active intense accretion ceased. The lower luminosity boundary of the rectangle corresponds   
to the upper luminosity limit obtained from the SED, while the upper   
luminosity boundary corresponds to the luminosity, obtained from the   
optical light curves modeling. Given all uncertainties, the agreement   
with mass estimate obtained above seems to be rather good. But we note   
cautionary 
that the tracks were calculated assuming solar chemical composition of   
 accreted matter, while  real chemical composition of the {\sl hot}   
component is not known and that the structure of its envelope, which   
defines the rate of cooling, may be different from the one of a   
steady-burning object.

For the readers convenience we summarize parameters of the binary estimated by variety of methods in Table\,\ref{tab:TSparameters}.

 \begin{table*}
\caption{Summary of parameters of \TS.}  
\label{tab:TSparameters} 
\scalebox{0.8}{
\begin{tabular}{lcccccccc}       
\hline  
  & T$_{\mathrm{cool}}$   & M$_{\mathrm{cool}}$  & R$_{\mathrm{cool}}$ & log\,$g_{\mathrm{cool}}$  &  T$_{\mathrm{hot}}$  & M$_{\mathrm{hot}} $ &  R$_{\mathrm{hot}}$  &used data \\
 Method  & $\times10e3$\,K  & \msun & \rsun &   &  $\times10e3$\,K  & \msun & \rsun &  \\
       \hline 
%     DFT  &   0\fd163508 &      &    &     &    &     &    & Calar-Alto, SPM, XMM, HST  \\ 
    FITSB2        &   $60\pm5$ &  &  & $5.17\pm0.07$  & &  &  &  GEMINI, CFHT\\
    SED 2BB fit   &  $55\pm7$ &  &  0.43\tablenotemark{*}&                              &  152--175  &&0.04--0.12 & SDSS, HST, FUSE, XMM\\
 Evolution       &  & $0.53\pm0.01$  &   &                &       &    & &  Weiss \& Ferguson (2009)  \\
  NIGHTFALL & 57 \tablenotemark{*} &  0.54 \tablenotemark{*}& $0.43\pm0.02$ & $5.03\pm0.02$& 160--200 & 0.71--0.93& 0.1--0.13 &  Calar-Alto, GEMINI, CFHT\\
  TMAP & $55\pm5$ & &  &  $4.9\pm0.5$ &&&& GEMINI, HST \\
\hline
\end{tabular}
}
\tablenotetext{*}{Fixed parameters in corresponding fits}
\end{table*}

\section {Evolutionary considerations.}
\label{sec:evol}

\subsection{Formation of \TS}
\label{sec:form}

Below, we  present examples of evolutionary scenarios that can result in the formation of a system
reasonably similar to \TS.  These scenarios envision the formation of a massive white dwarf 
that cools for a long time but is reheated by compression due to accretion and nuclear burning 
of material captured from the stellar wind of companion. 
%that is later heated by nuclear burning of material accreted from the wind of a giant companion,
The core of the latter is currently the less massive component of the binary nucleus of \TS.
In the analysis of the origin of \TS, one has to take into account its location in the Galactic halo, where star formation ceased about 10\,Gyr ago \citep[e. g.][]{2009ApJ...694.1498M}.  This sets an upper limit of $<1$\,\msun\ for the mass of the progenitor of the {\sl cool} component.

The presence of 
%two white dwarfs 
an old white dwarf and a nascent white dwarf
in a system with an orbital period of 3.92 hr only implies 
that previous evolution of the system involved common envelope(s).  
First,  let us assume that both components of the 
%double-degenerate 
nucleus of \TS\ had AGB precursors.
This suggests the following scenario{\footnote{Below, we use the terms ``primary'' and ``secondary'' for components that were, respectively, more and less massive at ZAMS.  As we will show further, ``primary'' is precursor of the {\sl hot} component, while ``secondary'' is precursor of the {\sl cool} component.}}. The initial masses of the components 
are  significantly different.
The initial system is so wide that the primary evolves to the AGB
unaffected by the presence of the companion. In the FGB\footnote{The first giant  branch, in which the helium nucleus is formed but not burning yet.} and AGB the system might
manifest itself as a symbiotic system with AGB and main-sequence components \citep{1984ApJ...279..252K}. 
Owing to the wind mass loss from the system, the separation of the components increases. Both the radius of the primary and the radius of its Roche lobe increase, with the radius of the primary growing faster until the primary overflows its Roche lobe 
(RLOF) close to the tip of the AGB.  
Both because the primary at this time has a deep  convective envelope and the mass ratio of the components 
is high, the dynamical mass loss is unavoidable \citep{1987ApJ...318..794H}  and the shedding of the envelope results in formation of a common envelope (CE) and a reduction of the separation of the components due to angular momentum loss in CE.
What remains of the primary after this episode is the more massive component of the core of \TS.
The system remains wide enough so that other component  may evolve to become a giant star too
and experience RLOF close to the tip of the AGB, forming the current  {\sl cool} component.  The matter ejected during the second CE episode is now observed as a planetary nebula.  

We now present  numerical estimates that argue in favor of  the feasibility  of a scenario such as that just described.   In our evolutionary simulations, we use the ``rapid evolutionary code'' SSE \citep{hpt00} based on the analytical fits to detailed grids of full stellar models.  We use the SSE code because detailed evolutionary tracks for low-metallicity stars with $M < 1$\,\msun\ have yet to be computed.  Comparison with data for more massive stars \citep[e. g. ][]{2009A&A...508.1343W} shows that the initial-final mass relations used by us agree with the results of sophisticated, full evolutionary computations to within $\simeq 10$ per cent  and, hence, qualitatively, the resulting scenario must be robust.  We note also, that the results of evolutionary calculations depend heavily on the opacities used for the models.  

\citet{paperI} estimate that the metallicity of \TS\ ranges from 1/12 to 1/30 of the solar value 
\citep[taken as $Z_\odot$=0.014;][]{2009arXiv0901.1149L}.
For our calculations we accepted Z=0.001 as a proxy to the metallicity of \TS, 
since our goal is to demonstrate the possibility of forming a system similar to \TS, rather than attempt to reproduce precise values for parameters that are still quite uncertain.
% Parenthetically we may note that at the end of AGB stage
%low-mass stellar models obtained with SSE for Z=0.001 and 0.0005 are virtually similar.

The second CE episode, which produced the current  {\sl cool}  component, followed RLOF 
by its precursor close to the tip of AGB.   Using SSE, we find that, for Z=0.001, stars  with a ZAMS mass exceeding 0.89\,\msun\ evolve to the tip of AGB in less than 10 Gyr.  At the tip of the AGB, a star with  $M_{\rm ZAMS}=0.89$\,\msun\ 
has a  mass of 0.60 \msun\ and a  core mass of 0.54 \msun, which is coincidentally similar to the estimated mass of  the {\sl cool}  component in \TS. Motivated by Fig.~\ref{mtot_chi2nightfall}, we adopt a total system mass of 1.39\,\msun.  Given a mass of 0.54\,\msun\ for the 
the  {\sl cool} component,  the mass of {\sl hot} component is then $\approx 0.85$\,\msun, corresponding to an initial mass of 2.5\,\msun.

At the tip of AGB, the precursor of the {\sl cool} component had a radius $R\approx 140$\,\rsun. Using the formula from \citet{egg83} for the dimensionless radius of the Roche lobe $r_L$, we estimate that the pre-CE separation of components was about $400$ \rsun.
The variation of the separation of components in CEs may be described by the formula suggested by 
\citet{web84}:
\begin{equation}
\label{eq:web84}
\frac{a_f}{a_0} = \frac{M_c}{M_2}\left[1+\frac{2}{\al r_{2,L}}\frac{M_2-M_c}{M_1}\right]^{-1},
\end{equation}

\noindent where $M_2$ and $M_c$ are initial and final masses of mass-losing component  (the donor), 
$M_1$ is the mass of companion, \al\ is the product of efficiency of common envelope expulsion \ace\ and the structural parameter $\lambda$\ which characterizes binding energy of the donor envelope.  $r_{2,L}$ is the fractional Roche lobe radius of the donor.
The reduction of the separation from  $a_0 \approx 400$ \rsun\ to the current $a_f \approx 1.3$ \rsun\ is possible if $\al \approx 0.0015$, i.e., is extremely low.  

In the stage preceding the common envelope, the system contained an AGB star and a massive ({\sl hot}) companion accreting from the wind.  In this stage, the system could be identified with a symbiotic star \citep{ty76,1984ApJ...279..252K,1995ApJ...447..656Y,2006MNRAS.372.1389L}. 
% Accretion could result in transient or persistent emission of 
%supersoft X-rays \citep{hbnr92,1995ASSL..205..453T,ylttf96} and heating of the white dwarf.
Accretion reheated white dwarf and resulted initially in unstable and later in stable hydrogen burning 
at the surface of white dwarf. Energy release by nuclear burning also contributed to the heating of white dwarf. 

\begin{figure}[ht]
\includegraphics[width=8.2cm,bb = 0 18 1030 790, clip=]{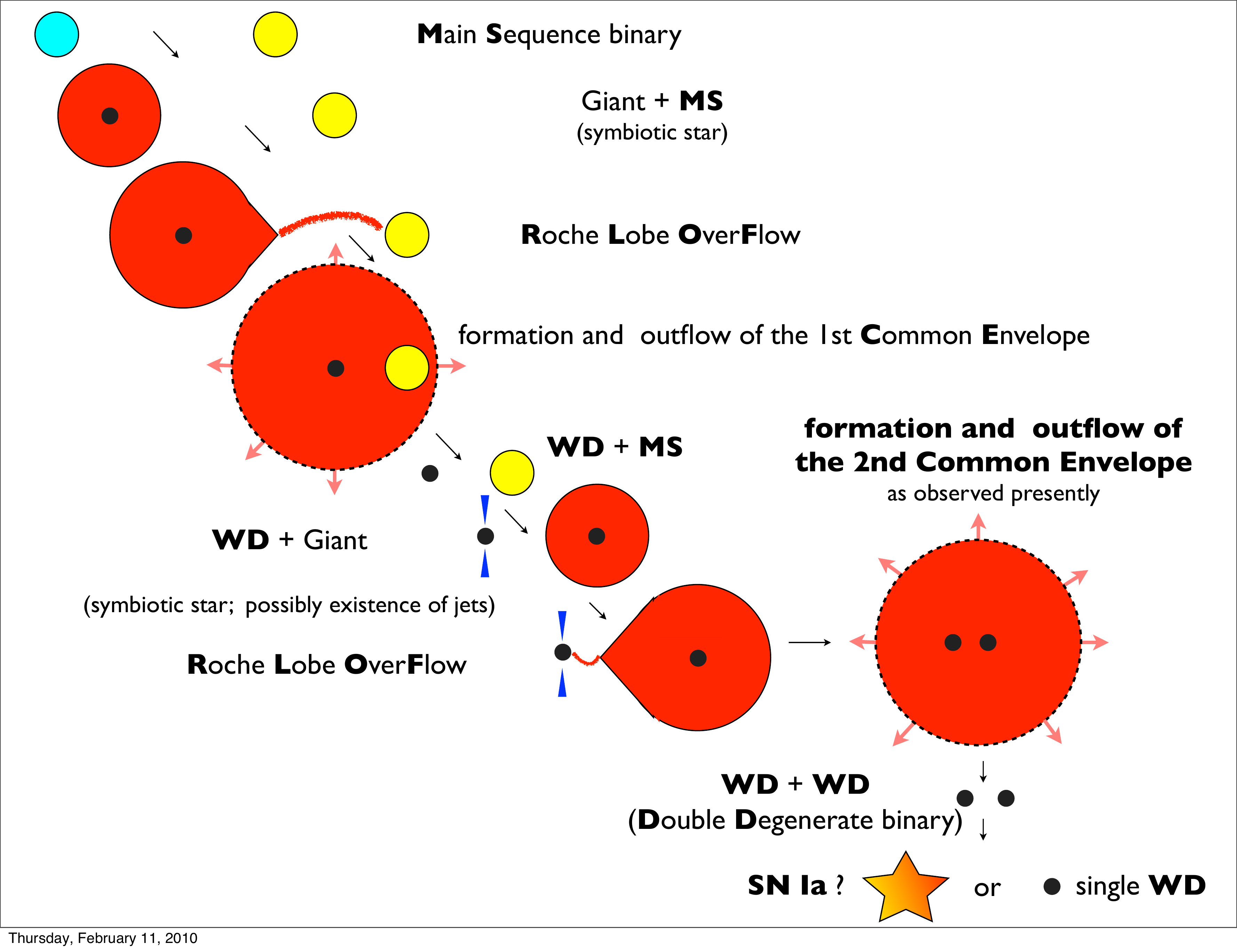}
\caption{Cartoon depicting the evolution of \TS. }
\label{scheme}
\end{figure}

  It is plausible that currently {\sl hot} component still burns remainders of hydrogen accreted in this stage. During the symbiotic stage, the precursor of the {\sl cool} component lost about 0.29\,\msun\ via a wind. Accretion from the wind in symbiotic systems is inefficient \citep[$\sim 10\%$][]{2009ApJ...700.1148D} and we may safely assume that all mass lost by the donor was lost from the system taking away specific angular momentum  of the donor (``Jeans mode of mass ejection'') and that the mass of the {\sl hot} component did not change.  Jeans mode of mass ejection has an invariant $a \times (M_1+M_2)$ and, hence, the separation of the components in the beginning of the symbiotic stage was 320\,\rsun.  This separation, 320\,\rsun, is also the separation of components after the first CE stage, which aborted 
the  ascend of AGB by the initially more massive component close to the tip of the AGB. Before the first CE stage, the mass of 
the star decreased via wind mass-loss from 2.5\,\msun\ to 1.29\,\msun.  Its radius at the tip of the AGB was 495 \rsun. Like for the second RLOF episode, from the condition of RLOF we may estimate that the separation of  the components at the beginning of the first RLOF was 1200\,\rsun.  The reduction of the separation in the first CE phase from 1200\,\rsun\ to 320\,\rsun\ implies $\al \simeq 1.5$.  The first CE episode could also have been preceded by a symbiotic stage. We again assume that all mass lost by the donor was lost from the system via the Jeans mode of mass ejection. We then estimate the initial separation of components as close to 770\,\rsun. We neglect the wind mass loss during the first red giant stage, which is only several 0.01\,\msun\ for $M_0\simeq2.5$\,\msun.  The numerical data is summarized in the upper part of Table \ref{tab:scenario} (scenario I)  and is presented in a form of a cartoon in Fig.\,\ref{scheme}.

\begin{table*}[ht!]
\caption[]{Numerical evolutionary scenarios for \TS. The upper part of the Table shows scenario (I) based on 
energy balance formalism of \citet{web84}, while lower part of the Table presents scenario (II) based on 
angular momentum balance formalism of \citet{2000A&A...360.1011N}} 
\begin{tabular}{llll}\tableline\tableline \\
$M_1$ &$M_2$   &$a$  & Comment \\ 
\msun &\msun   &\rsun  &  \\ 
%&&&\\
\hline
&&&\\
2.5 & 0.89  & 770 & ZAMS \\
1.29 & 0.89 & 1200  & The end of AGB ascend by the primary, beginning of the first RLOF (CE) \\
0.86 & 0.89  & 320 &  The end of the 
first CE, $\ace\approx1.5$, formation of the first WD, \\
&&&beginning of the symbiotic stage \\
0.86 & 0.60  & 400 &  The end of the symbiotic stage; RLOF (CE); ejection of PN, $\al \sim 0.001$ \\
0.86  &  0.54 &  1.3 & Present state \\ 
&&&\\
\tableline
&&&\\
%&&&\\
\tableline
&&&\\
5.0 & 0.89 & 150 & ZAMS\\
5.0 & 0.89 & 150 & The end of FGB ascend by the primary,  beginning of the first RLOF (CE)\\
0.87 & 0.89 & 240 & The end of the first CE, $\gamma\approx1.2$, \\
&&&formation of He-star evolving into first WD, beginning of the symbiotic stage \\
0.87 & 0.73 & 260 & The end of secondary evolution in E-AGB; RLOF (CE); ejection of PN, $\al \sim 0.01$ \\ 
0.87  &  0.53 &  1.3 & Present state \\
\tableline\tableline
\end{tabular}
\label{tab:scenario} 
\end{table*}

An apparent problem with the suggested scenario is the large difference in \al\ for the common envelope stages.  While $\al \sim 1$ is typical for WD+MS stars with periods below about 10 days, which are supposed to form via one common envelope stage and implies that the energy spent on the expulsion of the common envelope is comparable to the orbital energy of the initial binary
 \citep{nelemans_tout05}, 
$\al \sim 0.001$ during the second CE episode appears atypically low. 
However, common envelopes  remain virtually \textit{terra incognita} in stellar evolution and we 
cannot exclude a significant  difference in the interaction of the AGB star envelope with a MS or 
a WD companion, which differ in structure and, most importantly, by two orders of magnitude in radius
 (whereas the drag force is $\propto R^2$).

An alternative scenario for \TS\ assumes that present {\sl cool} component had a 
precursor with ZAMS mass of 0.89\,\msun,
while the {\sl hot} component descended from a helium star which was formed
by RLOF close to the tip of FGB.  
For instance, a 5\,\msun\ star has a maximum  He-core mass of 0.87\,\msun\ which, presumably, evolves into a CO WD of the same mass\footnote{We set the mass of WD equal to the mass of its He-star precursor, thus implicitly neglecting the possibility of reexpansion of the He-star after exhaustion of helium in its core. Such 
an expansion with formation of a shallow CE and almost negligible mass loss was discovered by 
\citet{it85} for solar metallicity stars, but its possibility was never explored for non-solar metallicities.}.
Thus, the initial system could contain  a 5\,\msun\ component and a 0.89\,\msun\ component
and after the 1st CE to become  a (0.87+0.89)\,\msun\ system. If 5\,\msun\ star filled Roche lobe close 
to the tip of FGB when it radius was close to 80\,\rsun, prior to RLOF separation of components 
have had to be close to 150\,\rsun. 

If the precursor of the {\sl cool} component was an AGB star, the smallest radius with which it could
overfill its Roche lobe at E-AGB was $\approx 100$\,\rsun.
At this moment the total mass of the star was 0.73\,\msun, the mass of the core -- 0.53\,\msun, and  
from the RLOF condition we obtain that the separation of the stars was $\approx 260$\,\rsun.
In the second CE the separation decreased from 260\,\rsun\ to 1.5\,\rsun\ by ejection of 
0.2\,\msun. This is possible if $\al\approx 0.01$, i.e.  an order of magnitude larger than in the first scenario.

If we account for Jeans-mode mass loss  by the precursor of the {\sl cool} component
we obtain that, after the first CE, the separation of components was
close to 240\,\rsun. Thus we arrive to an apparent controversy: in the suggested scenario,
in the first CE episode, the separation of the components had to {\sl increase} from 150 to 240\,\rsun!   

However, it was noticed by \citet{2000A&A...360.1011N} and later confirmed by 
\citet{nelemans_tout05} that using Eq.~(\ref{eq:web84}) for the description of the outcome of unstable mass exchange between a giant and a MS-star 
often does not allow to reproduce well measured parameters of many post-CE binaries. As an alternative, \citet{2000A&A...360.1011N} suggested to estimate the post-CE 
separations of components using an equation for angular momentum balance:
\begin{equation}
J_{\rm i}-J_{\rm f}=\gamma J_{\rm i} \frac{\Delta M}{M_{\rm tot}}.
\label{eq:mom}
\end{equation}  
Here $J$ is the orbital angular momentum, subscripts ${\rm i}$ and ${\rm f}$ denote the  initial and final values of the momentum, $\Delta M$ is the mass lost from the system (the envelope of the donor), and $M_{\rm tot}$
is total initial mass of the system. Thus, a
single parameter $\gamma$ describes the fraction of initial specific
orbital angular momentum
of the binary taken away by 
outflowing matter. This ``$\gamma$-formalism" leads to:
\begin{equation}
\label{eq:ai_af_AM}
\frac{a_{\rm f}}{a_{\rm 0}} = \left(\frac{M_1}{M_{\rm c}}\right)^2
 \left(\frac{M_{\rm c} + M_2}{M_{\rm tot}}\right) 
\left(1 -\gamma \frac{\Delta M}{M_{\rm tot}}\right)^2.
\end{equation} 
Here $M_{\rm c}$ is the mass of the core of the mass-losing component. 
An increase of  the separation during the  CE from 150\,\rsun\ to 240\,\rsun\ is possible if $\gamma\approx 1.2$. 
Rather similar combinations of $\gamma$ for the first CE and $\al$ for the second one
were found for  some systems 
studied by \citet[][ see their Figs. 1 and 5]{nelemans_tout05}.

\subsection{Common Envelope Remnant vs. Single Star Evolution through post-
AGB phase.}

\begin{figure}[ht]
\includegraphics[width=8.cm,bb = 0 0 400 400, clip=]{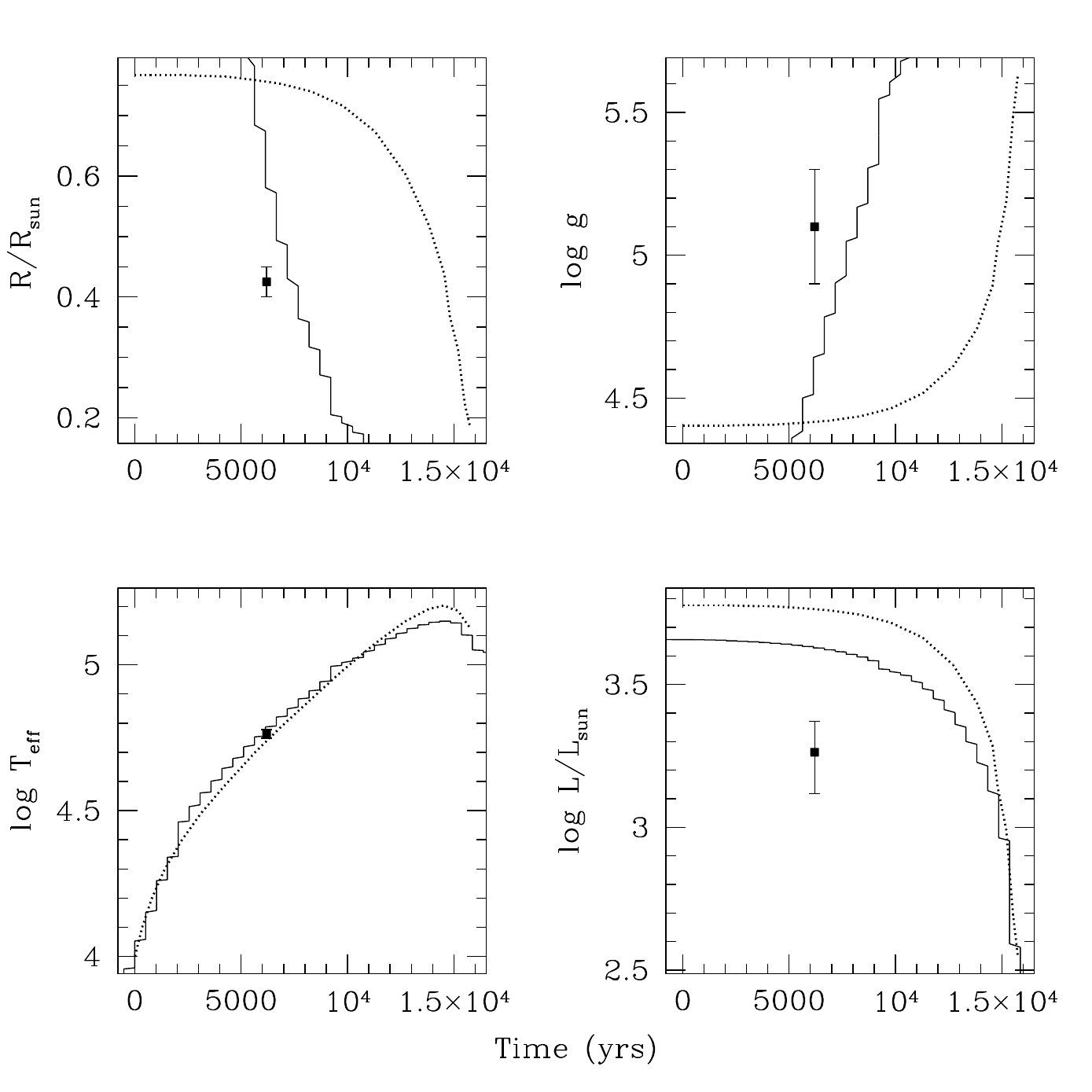}
\caption{Various 
%quantities of the AGB evolution models 
parameters of the models of post-AGB stars vs. time. Solid line is for the $M=0.547$\,\msun, Z=0.0005
model \citet{2009A&A...508.1343W},
% and 
dotted line is for the $M=0.623$\,\msun, Z=0.001model \citet{1994ApJS...92..125V}.
\TS\ is marked by  squares with error bars  at age of ($\sim6000$\,yr), 
appropriate to its effective temperature. 
%where it crosses with evolutionary tracks (bottom left panel).  The tracks are for single post-AGB stars of 1.0\,\msun\ and the lowest available Z from the models (0.0005 \& 0.001, respectively).
%There is a large discrepancy of deduced parameters for \TS\ (R$_{\rm {cool}}$ or log $g$) with  the \citet%{1994ApJS...92..125V}'s model for that age. 
}
\label{evol_sbs}
\end{figure}

%The low values of \al\ deduced in previous section testify that the expulsion of the envelope by the {\sl cool} component was rather forceful than  simple detachment of shell in a single star. 
Ejection of a common envelope definitely differs from formation of a planetary nebula by the usually assumed 
superwind mechanism.  In that context, it is interesting to compare parameters of the {\sl cool} star deduced here with the evolutionary models 
%existing today. 
for post-AGB stars.
Given that the {\sl cool} component of TS01 nearly fills its Roche lobe and 
that it is currently contracting, it has only recently terminated 
%a 
the phase of common envelope evolution.  
%Its 
%atmosphere,
The structure and mass of its envelope might be very different from that of a single star passing through the early   
epochs  of 
%its 
planetary nebula nucleus stage.  
%As a result, it might be expected to be 
%considerably smaller than a single star at the same evolutionary phase. 
%The models for stars of low initial masses  with low metallicities have made considerable progress recently. 
Here we compare the derived parameters of the {\sl cool} 
%our observations 
component of \TS\ with  two sets of models of remnants   
%Indeed, the theoretical models 
of single stars with initial masses of 
1.0\,\msun (lower progenitor masses are not available in the literature).
Based on estimated abundances, we selected the models
$M=0.623$\,\msun, Z=0.001 \citet{1994ApJS...92..125V} and
$M=0.547$\,\msun, Z=0.0005\  \citet{2009A&A...508.1343W}. 
These models agree well regarding the time-dependence of heating of the core of a post-AGB star (lower right panel of Fig.\,\ref{evol_sbs}).
\TS\ 
%crosses these evolutionary tracks 
has $T_{\rm eff}$\ similar to them  at the age of $\approx6000$ yr. This 
%is also 
age estimate is in a good agreement with the age deduced from the expansion velocity and distance to the planetary nebula \citep{paperI}. 
%However the models in the other panels of Fig.\,\ref{evol_sbs} are largely discrepant.  The position of the \TS\ on the rest of  panels corresponds to the age determined from heating time (the bottom right panel) and corresponding parameters (R,  $log$\,g and $log$\,L) as determined in this paper. The shaded boxes across the upper panels indicate errors of determination of corresponding parameters.  As one can see, the single post-AGB model stars  have larger radii and 
%smaller surface gravities than we observe.  The newer models of \citep{2009arXiv0903.2155W} seem to better reflect the position of \TS\ on its evolutionary path, but actually they coincide for an object of $\sim6000$  year of age. According to both models a single post-AGB star of that age should have R about 0.75\,\rsun\ and $log {\mathrm g}\approx4.5$.  

Compared to the most modern and the closest in mass 
model  of \citet{2009A&A...508.1343W}, the nucleus of \TS\ is slightly more compact
(by 0.05 dex) and significantly (by more than 0.3 dex) less luminous. Since the main source of luminosity of post-AGB
stars is hydrogen burning, this may mean that common envelope remnants may have less massive H/He envelopes 
around degenerate cores than their 
post-AGB counterparts.

This comparison clearly indicates that 
%evolution post-AGB evolution of 
evolution in common envelopes might alter evolution of stars  in  close binary systems compared to single stars  
%might be altered 
and a more complete analysis of TS01 is warranted than can be  made with single star models.

\subsection{\TS\ and SNe Ia}
\label{sec:snia}

The evolutionary path  suggested  for \TS\ includes a stage of a symbiotic star which is 
considered as one of the routes to \sna\ \citep[e. g.][]{ty76,it84a,1992ApJ...397L..87M}. 
However, conditions in symbiotic systems are not favorable for an efficient accumulation of matter by the white dwarf components.
Accretion from the wind typically allows only several per cent of the mass lost by the donor to be accreted. In the numerical scenario above, the  maximum mass-loss rate by the progenitor of the {\sl cool}  component estimated by 
means of SSE is close to $2\times10^{-7}$\,\myr. This means that for about 10\,Gyr the white-dwarf ({\sl hot}) component
stays in the regime of unstable thermonuclear burning (of Novae eruptions)  
\citep{nom82a} and instead of accumulating mass it may erode.
Conditions for accretion ``improve'' 
if the accretor is located in the zone of acceleration of the stellar wind, which requires the proximity of the donor surface to the Roche lobe \citep{1995ApJ...447..656Y}, or if the stellar wind is pumped close to the Roche lobe by pulsations and still remains slow \citep{2007BaltA..16...26P}.
Then accretion efficiency may become close to 100\%. 
Using \citet{1944MNRAS.104..273B}  formalism for wind accretion,
accounting for possible location of accretor in the wind acceleration zone and taking accretion rate limits for stable
hydrogen burning after \citet{nom82a}, we estimate that the system could accrete steady for the 
last several $100\,000$\ yr prior to CE and accumulate only several 0.01\,\msun. In a more general context, the circumstances listed above prevent symbiotic stars from being efficient progenitors of \sna\ and the estimated rate of occurrence of \sna\ in these systems is only $\sim 10^{-6}$\,\pyr\ on a Galactic scale \citep[e. g.][]{2005ASSL..332..163Y}. 

For a fraction of time between Novae eruptions and in steady-burning regime
the system can manifest itself as a supersoft X-ray source
\citep[e. g][]{hbnr92,1995ASSL..205..453T,ylttf96}.

Note that, during the stage of accretion onto the current {\sl hot} component, the matter could inflow onto the equatorial regions of the dwarf while it outflows from the polar regions.    The ``bars''  seen in \TS's\ nebular shell (Stasi\'nska et al. 2009)  
may be the remnants of jets that  once existed in the system.

\begin{figure}[ht!]
\includegraphics[width=7.8cm,bb = 130 390 470 740, clip=]{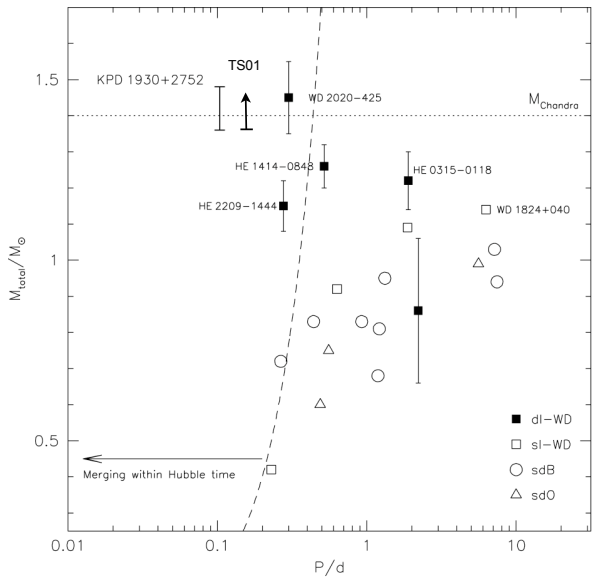}
\caption[]{Positions of the known detached double-degenerate  systems (DD)
and subdwarf+white dwarf systems which will turn into double degenerates.
d1 -- DD with two visible spectra, s1 -- DD with one visible spectrum. Updated figure from 
\citet{2007A&A...464..299G}. The position of \TS\ is marked by an arrow, indicating the tendency of better fit to the observed data with increase of total mass of the binary. The minimum value $M_{\mathrm {tot}} \ge 1.35$\msun\ corresponds to the leveling of   $\chi^2$  slope in Fig.\,\ref{mtot_chi2nightfall}. We consider the values at plateau in the $\chi^2 - M_{\mathrm {tot}} $  dependance as a range of  best solutions.}
\label{fig:spy}.
\end{figure}

Figure~\ref{fig:spy} shows the positions of  double-degenerate systems with known 
parameters in  the  $M_{\rm{tot}} - P_{\rm{orb}}$ plane.  As well, positions of several sdB stars with white 
dwarf companions are shown. The latter systems will turn into double-degenerates after 
completion of helium burning in sdB stars. Thus, its short orbital period of 3.92 hr and its  total mass close to the Chandrasekhar mass makes \TS\ very promising candidate progenitor for a \sna\ in the double-degenerate scenario for these events \citep{ty81,it84a,web84}. For instance, the merger of components will occur in $\approx$660~Myr 
%for 
in the first evolutionary scenario  suggested above
%solution 
and in  
$\approx$1.2~Gyr 
%for the second solution  
in the second scenario. 
The only ``competitor'' to \TS\ is sdB+WD system KPD 1930+2752 with orbital period 2.28~hr,
$M_{\rm{sdB}}$ = 0.45 -- 0.52\,\msun, $M_{\rm{tot}}$ = 1.36 -- 1.48\,\msun\
\citep{2007A&A...464..299G}.
In the latter system, subdwarf star will turn into a WD in $\approx$(220 -- 140)~Myr,   
see \citet{2008AstL...34..620Y} for estimates of lifetime of sdB stars.
It will take two WD   several tens of Myr more to merge. 
Favourable conditions for central carbon ignition may come to fruition just in systems with 
low mass ratios of components \citep{2007MNRAS.380..933Y}, like \TS\ and 
KPD 1930+2752.

\section{Conclusions.}
After  a decade of intense study, we have achieved a good  understanding of an object whose discovery spectrum was misidentified  and incomprehensible  in 1997. Since then, the object has been observed at practically all wavelengths  with the  help of the most advanced instruments.  This paper accompanies \citet{paperI}, which focuses upon the chemical composition and ionization state of \TS's nebular shell. Here, we  focus on the nature  of the close binary nucleus of the PN.    

%The binary is one of the shortest period systems  among the PNe with binary cores, with an orbital period of 3.924 hours.

\TS\ is one of the shortest period systems among the 
double-degenerate or pre-double-degenerate systems, with an orbital period of 3.924 hours
%Unlike any other binary central star  of a PN, it is a double-degenerate system.
This fact would not have caused confusion if the older of the 
%two white dwarfs  
components were  significantly cooler than the core of the star that most recently ejected its envelope to form the current PN.  However, observations and analysis clearly demonstrate that \TS's nucleus is comprised of two compact stars, both extremely hot and thus, both being sources of ionization for the nebula.  This unusual phenomenon created  confusion and misinterpretation of the object in the past. Nevertheless, the correct understanding of the ionization source does not change the essence of those previous interpretations. \TS\ remains a PN with a record low oxygen abundance \citep{paperI}.

%According to our scenario, the object evolved from the wide binary through two common envelope episodes. In the current stage we are observing the second common envelope, which  formed the PN.  The core of the envelope-shedding post-AGB star is in the process of contraction and heating up. As for  the lower mass component, at the present time, it  nearly fills its Roche lobe and has an ellipsoidal shape.  Recently, the more massive companion, which became a white dwarf earlier, underwent a period during which it accreted mass at a high rate and, since then, has maintained steady nuclear burning at its surface, resulting in a change in its structure and its heating up to the temperatures typical of supersoft X-ray sources.   It is believed that a fraction of super soft X-ray sources are symbiotic stars, and \TS\ has just emerged from such a   state, becoming one of the softest X-ray sources ever, similar to  Lin 358  \citep{2007ApJ...661.1105O}.  

According to our scenario, TS\,01\ evolved through two common envelope episodes. In the current stage we are observing the remainders of the second common envelope as a PN.  The core of the 
envelope-shedding post-AGB star is in the process of contraction and heating up. At the present time, it  nearly fills its Roche lobe and has an ellipsoidal shape.  Before the last CE episode, 
the more massive component, which became a white dwarf earlier, underwent a period during which it accreted mass at a high rate and burned hydrogen steady. Since then it stays
close to the temperatures range typical for  supersoft X-ray sources.  Its properties make \TS\ one of the softest X-ray sources ever, similar to  Lin 358  \citep{2007ApJ...661.1105O}.

The parameters of the binary system were deduced using a wealth of information and via three independent routes.  Although, each of these methods requires its own assumptions and each alone produces ambiguous  results, in combination, they converge to values with unusual precision.  Using the spectral energy distribution, from the far infrared to X-rays, the light and radial velocity curves, and by fitting atmospheric models to the stellar absorption features of the {\sl cool}  component, we find  that the {\sl cool} component has a mass of $0.54\pm0.2$\,\msun, an average T$_{\mathrm {eff}}$ of $58\,000\pm3\,000$\,K, a mean radius of $0.43\pm0.3$\,\rsun, and $\log g=5.0\pm0.3$. The {\sl cool} component nearly fills its Roche lobe.  The temperature and gravity over the surface of the {\sl cool} component are not homogeneous.  
 
The chemical composition of the {\sl cool} component from atmosphere model fitting was determined as: 12 + log He/H = 10.95 and 12 + log C/H = 7.20, with an uncertainty of about 0.3\,dex, and upper limits  12 + log N/H $<$ 6.92 and 12 + log O/H $<$ 6.80. Overall, the agreement with the abundances found in the nebula by \citet{paperI} is very good, except for the carbon abundance, which is found to be higher in the nebula for a reason yet not understood. 

The parameters for the {\sl hot} component are less certain. It is fairly clear that the spectral energy distributions of real stars at such high temperatures depart from that of a black body. The range of temperatures that we determined for the {\sl hot}  component spans 160--200\,kK.  It seems that  the real object acts like a 180-200\,kK blackbody in the X-ray range but appears as a 160\,kK blackbody in the UV/optical range. Uncertainty in its temperature leads to uncertainty in its size, but it is obvious from our calculations that the {\sl hot} component  is  larger than normal for a white dwarf, R$_{\mathrm hot} > 0.1$\,\rsun, and is probably bloated as a result of intense accretion in the recent past.  However, we have indirect information on the hot component through photoionization modeling by  reproducing the intensities of the lines emitted by the nebula \citep{paperI}. We estimate the distance to the object as $\sim 21$ kpc, and our most reasonable luminosity estimate for the X-ray component is $\sim 10^4$\,\lsun, appropriate for a supersoft X-ray source.

The total mass of the binary is very close to Chandrasekhar limit.  This makes \TS\ one of the best of the known candidates for the progenitor of a type Ia supernova.

\acknowledgments
We appreciate assistance of J.Greiner helping to check the ROSAT data. R.Wichmann kindly clarified our inquires about {\sl Nightfall}.
GT acknowledges UC-MEXUS grant allowing his visit to UCSD and hospitality of Center of Astrophysics and Space Science during the visit.
JT and GT were supported by a grant NNX07AQ12G associated with the  XMM-{\sl Newton} observations. GT acknowledges CONACyT financial support from project 45847 and PAPIIT 101506 .
LRY is supported by RFBR
grant 07-02-00454 and Presidium of the Russian Academy of Sciences
Program ``Origin, Evolution and Structure of the Universe Objects''.
VS thanks DFG for financial support (grant SFB/Transregio~7 
"Gravitational Wave Astronomy"), 
and for partial support President's programme 
for support of leading science schools 
(grant \hbox{NSh-4224.2008.2}), and 
 RBRF(grant \hbox{09-02-97013-p-povolzh'e-a}).
 T.R\@. was supported by the German Astrophysical Virtual 
Observatory (GAVO) project 
of the German Federal Ministry of Education and Research (BMBF) under 
grant 05\,AC6VTB and 
by the German Aerospace Center (DLR) under grant 05\,OR\,0806. 
MGR gratefully acknowledges financial support throughout this project from CONACyT grants 43121, 49447, and 82066 and DGAPA-UNAM grants 108406, 108506, 112103, and 116908.

{\it Facilities:} \facility{XMM}, \facility{HST (STIS)}, \facility{Gemini},  \facility{CFHT},  \facility{SDSS}, \facility{Calar-Alto(BUSCA)},  \facility{OAN SPM}.

\bibliography{sbs_2009}

\end{document}